\documentclass[twocolumn,trackchanges]{aastex701}

\usepackage{CJKutf8}

\usepackage{graphicx}
\usepackage[caption=false]{subfig}
\usepackage{tabularx}
\usepackage{capt-of}
\usepackage{hhline}
\usepackage{lineno}
\usepackage{amsmath}	
\usepackage{amssymb}	
\usepackage{textcomp}
\usepackage{upgreek}
\usepackage{listings}
\usepackage{pifont} 

\begin{document}

\title{The Road to Normalcy: Environment-Driven Evolutionary Pathways for Primordial Black Holes}

\author[orcid=0000-0003-1541-177X,gname=Saiyang,sname=Zhang]{Saiyang Zhang\begin{CJK*}{UTF8}{bsmi}（張賽暘）\end{CJK*}}
\affiliation{Cosmic Frontier Center, The University of Texas at Austin, Austin, TX 78712, USA}
\affiliation{Department of Physics, University of Texas at Austin, Austin, TX 78712, USA}
\affiliation{Weinberg Institute for Theoretical Physics, Texas Center for Cosmology and Astroparticle Physics, \\ University of Texas at Austin, Austin, TX 78712, USA}
\email[show]{szhangphys@utexas.edu}

\author[orcid=0000-0002-6038-5016,gname=Junehyoung,sname=Jeon]{Junehyoung Jeon}
\affiliation{Department of Astronomy, University of Texas at Austin, Austin, TX 78712, USA}
\affiliation{Cosmic Frontier Center, The University of Texas at Austin, Austin, TX 78712, USA}
\email{junehyoungjeon@utexas.edu}

\author[orcid=0000-0002-4966-7450,gname=Boyuan, sname=Liu]{Boyuan Liu\begin{CJK*}{UTF8}{bsmi}(劉博遠)\end{CJK*}} 
\affiliation{Institute of Astronomy, University of Cambridge, Madingley Road, Cambridge, CB3 0HA, UK}
\affiliation{Universit\"at Heidelberg, Zentrum fur Astronomie, Institut f\"ur Theoretische Astrophysik, D-69120 Heidelberg, Germany}
\email[show]{boyuan.liu@uni-heidelberg.de}

\author[orcid=0000-0003-0212-2979,gname=Volker,sname=Bromm]{Volker Bromm}
\affiliation{Weinberg Institute for Theoretical Physics, Texas Center for Cosmology and Astroparticle Physics, \\ University of Texas at Austin, Austin, TX 78712, USA}
\affiliation{Department of Astronomy, University of Texas at Austin, Austin, TX 78712, USA}
\affiliation{Cosmic Frontier Center, The University of Texas at Austin, Austin, TX 78712, USA}
\email{vbromm@astro.as.utexas.edu}

\author[orcid=0000-0002-5554-8896,gname=Priyamvada,sname=Natarajan]{Priyamvada Natarajan}
\affiliation{Department of Astronomy, Yale University, New Haven, CT 06511, USA}
\affiliation{Department of Physics, Yale University, New Haven, CT 06520, USA}
\affiliation{Black Hole Initiative, Harvard University, 20 Garden Street, Cambridge, MA 02138, USA}
\email{priyamvada.natarajan@yale.edu}



\begin{abstract}
We investigate how cosmological environment regulates the evolution of primordial black hole (PBH) seeds in the early universe using a suite of hydrodynamical simulations. PBHs, formed during the radiation dominated epoch, offer a possible source of early BH seeds that does not require prior star formation.
We find that PBHs follow distinct evolutionary pathways depending on the gas supply and halo assembly history. 
In underdense regions, limited inflow suppresses both accretion and star formation, producing faint, metal-poor systems, while in overdense environments, sustained gas inflow drives rapid BH growth and the formation of compact, centrally concentrated stellar components. 
These differences lead to large variations in BH-to-stellar mass ratio, metallicity, and morphology, collectively bracketing a range of possible evolutionary pathways for PBH-seeded systems.
Some pathways pass through a phase with compact stellar structure and elevated BH-to-stellar mass ratio that resembles observed Little Red Dots, before evolving into more extended galaxy--AGN systems.
Our results suggest that both the initial seed properties and cosmological environment jointly shape early BH--galaxy co-evolution, while subsequent environmental regulation can erase the memory of the initial seeding channel.
\end{abstract}

\keywords{\uat{Dark matter}{353} --- \uat{Early universe}{435} --- 	\uat{Galaxy formation}{595} --- \uat{Population III stars}{1285} --- \uat{Supermassive black holes}{1663}}


\section{Introduction}

 Recent JWST observations have transformed our knowledge of galaxy and black-hole (BH) assembly in the first billion years of cosmic history \citep[e.g.,][]{AdamoRev2025}. Massive galaxies, luminous AGN candidates, compact red sources, and chemically young systems have now been identified at redshifts where standard models of early structure formation are still being actively tested \citep[e.g.,][]{Finkelstein2022, Vanzella_LAP12023,Labbe2023Natur.616..266L,Boylan2023,Natarajan:2023UHZ1, KokorevLRD2024ApJ...968...38K,Scholtz2024A&A,Taylor2025, MaiolinoHebe2026arXiv}. A recurring theme in these discoveries is not simply that individual objects appear extreme, but that they reveal a surprisingly diverse set of early BH--galaxy configurations. At the redshift frontier, systems such as GN-z11, UHZ-1, and other candidate galaxies hosting accreting BHs point to rapid BH growth already at $z\gtrsim10$ \citep[e.g.,][]{Bogdan:2023UHZ1, GHZ9Napolitano2025ApJ...989...75N, Chavez2025arXivGHZ2,NaiduMoMz142026OJAp,Fabian2026MNRAS.547ag379F}. At somewhat lower redshifts, compact red sources such as Little Red Dots (LRDs) often combine small sizes, red continua, strong emission lines, and evidence for AGN activity, suggesting a phase in which the nuclear component is difficult to disentangle from the compact stellar host \citep[e.g.,][]{Matthee2023,Leung2024:LRD, Taylor2025,KocevskiLRD2025ApJ...986..126K,LabbeLRD2025ApJ...978...92L, KorberLRD2026arXiv}. 
In the most extreme cases, the inferred BH-to-stellar mass ratios substantially exceed those observed in the local universe~\citep[e.g.,][]{MaiolinoBH2024A&A,TripleBHUbler2026,QSO1Direct2026Natur}, and some sources further show low metallicities or chemically young environments \citep[e.g.,][]{QSO1PristineMaiolino2025,Cliff2026MNRAS,MaiolinoHebe2026arXiv,UblerHebe2026arXiv}. These overlapping signatures suggest that JWST is probing not a single class of objects, but an ensemble of early BH growth modes and stages, ranging from high-redshift galaxies hosting AGN to compact, overmassive-BH systems and metal-poor nuclear sources. Together, these observations raise a broader question: how do early BHs and their baryonic environments co-evolve from diverse initial conditions and transient extreme phases toward the more familiar galaxy-AGN population observed at later times \citep{Kormendy2013GalaxySMBH,Smith_Rev2019,Inayoshi:2020}?

One possible route to early BH growth is provided by massive primordial black hole (PBH) seeds \citep[e.g.,][]{Cappelluti2022ApJ, Huang:2024aog, Ziparo2025JCAP...04..040Z, DeLuca2026PhRvL,  KashlinskyPBH2026MNRAS, Huang2026PhRvD.113b3519H, Allegrini2026arXiv260628296A}. Such seeds may form through several early-Universe mechanisms~\citep[reviewed in, e.g.,][]{Carr2020ARNPS..70..355C, EscrivaPBH2024}. Unlike stellar-remnant~\citep[e.g.,][]{Alexander2014Sci...345.1330A} or direct-collapse black hole (DCBH) seeding pathways, which are formed from the collapse of pristine gas \citep[e.g.,][]{BrommDCBH2003ApJ...596...34B, Begelman2006:DCBH, Lodato:2006DCBH}. While there exist multiple pathways to make DCBHs, a generic feature of all those models is the existence of a stage during which the BH is overmassive compared to the stellar component in the host galaxy \cite{Agarwal+2013, Natarajan+2017}.
PBHs can form before the onset of conventional star and galaxy formation and therefore need not be tied to the local thermal and chemical conditions, or the star-forming state of the gas. This makes PBHs a useful theoretical laboratory for studying how an initially massive compact object modifies subsequent structure and galaxy formation. Through their gravitational potential, PBHs can enhance local dark-matter clustering, deepen the central potential well, and promote gas accumulation \citep{Afshordi2003ApJ...594L..71A,Mack2007ApJ, Ricotti2007ApJI, Ricotti2008ApJII,ZhangPBH2024ApJ,Liu2025}. At the same time, accretion feedback can heat and ionize the surrounding gas, alter cooling, and regulate the onset of star formation \citep{LiuPBH2022MNRAS, Ziparo2022, Casanueva2024, ZhangPBH2024MNRAS,ZhangPBHGalaxy2025ApJ}. These coupled effects may also naturally produce phases in which the BH is overmassive relative to the stellar component, the host remains chemically young, or the emission appears compact and nuclear-dominated akin to what is predicted generically for all DCBH formation pathways \citep[e.g.,]{Dayal2024A&A...690A.182D,Dayal2026arXiv260420966D,DayalPBH2026A&A,Zhang2026PBHApJL}. Relatedly, \citet{Qin+2025} proposed a ``not-quite-primordial black hole'' (NQPBH) channel, in which moderately enhanced primordial density collapse into dark-matter halos at very high redshift ($(1+z) \gtrsim 200$) where the CMB suppresses molecular hydrogen formation, limiting gas cooling and fragmentation, thereby enabling direct collapse into massive black-hole seeds; we defer exploring the details of this distinct picture for future work.


In this work, we focus on the subsequent evolution of massive PBH seeds. The observable properties of PBH-seeded systems depend on both the initial seeds and the large-scale environments
in which they evolve~\citep[see e.g.,][]{Carr2018MNRAS.478.3756C,Wang2026arXiv260315736W}. 
In relatively isolated regions, limited gas inflow and feedback-driven gas removal can suppress sustained star formation and BH growth, leaving behind faint, metal-poor, or weakly evolved systems \citep[e.g.,][]{Dayal2024A&A...690A.182D,Zhang2026PBHApJL,Dayal2026arXiv260420966D}. 
In overdense regions, initial PBH clustering and mergers may accelerate the assembly of a more massive BH, whereas continued gas supply and hierarchical assembly can instead maintain both stellar growth and BH accretion, producing compact and luminous systems and accelerating the early formation of galaxies and AGNs \citep{Liu2022ApJL, DeLuca2023PhRvL.130q1401D, Matteri2025PBH,Liu2025}. Thus, the PBH-seeded systems may follow very different evolutionary pathways: they may remain an exotic, weakly fueled object for an extended period, temporarily appear as a compact BH-dominated source, or be incorporated into a growing galaxy and evolve toward a more conventional AGN-host system. This ``road to normalcy'' depends on whether subsequent galaxy assembly erases or preserves the initial PBH-driven signature.

This motivates a more detailed investigation of the astrophysical impact of PBHs in realistic cosmological environments. In previous studies, we have explored PBH effects on halo assembly, first star formation, and early BH--galaxy co-evolution \citep{LiuPBH2022MNRAS,ZhangPBH2024ApJ,ZhangPBH2024MNRAS,ZhangPBHGalaxy2025ApJ,ZhangDCBHPBH2025ApJ,Zhang2026PBHApJL}. Related studies of DCBH and early AGN formation have emphasized that compact BH-dominated systems may arise through multiple physical channels \citep{BrommDCBH2003ApJ...596...34B,JeonAGN2023,Hu2025,JeonSMBH2025ApJ...979..127J,JeonBHMF2025ApJ...988..110J,InayoshiLRD2024ApJ...973L..49I,InayoshiBinaryBH2026ApJ}. A controlled comparison between PBHs placed in different environments is therefore necessary to determine whether PBH signatures remain observable over cosmic time, and under what conditions they are erased by subsequent galaxy assembly.

We use cosmological simulations to follow PBH-seeded systems from the matter-dominated epoch at $z\sim3400$ to $z\sim3$ in both relatively isolated and overdense environments. Our fiducial PBH seeds are $10^5$ and $10^7\,M_\odot$, with mass scales motivated by scenarios involving the collapse of primordial fluctuations during early-Universe phase transitions \citep{Carr2021PDU....3100755C}; inhomogeneous Affleck--Dine baryogenesis, inflationary vacuum bubbles, or catalyzed dark-sector phase transitions \citep{Kawasaki2019PhRvD, Huang2024PhRvD, Kasai2025PhRvD, Guo2026arXiv}~\footnote{CMB $\mu$-distortion limits strongly constrain the abundance of massive PBHs formed from Gaussian primordial curvature perturbations~\citep[e.g.,][]{NakamaCMBDistortion2018PhRvD..97d3525N, CyrCMBDistortion2026arXiv}. Specific non-Gaussian statistics, alternative formation mechanisms, or non-Gaussianity combined with subsequent accretion can relax these constraints \citep[e.g.,][]{Balaji2026arXiv261112861B}, although their viability must be assessed within each model.}. These values bracket the BH masses inferred for the \textsc{glimpse} sources~\citep{GLIMPSEBH2026ApJ}, and for compact high-redshift systems such as the Cliff \citep[$z\simeq3.5$;][]{Cliff2026MNRAS}. Our simulations are performed as a controlled experiment, and do not assume a unique formation pathway or require a large cosmic PBH abundance.  As discussed in Appendix~\ref{Appendix:A}, seed number densities of $10^{-5}\,{\rm cMpc}^{-3}$ correspond to $f_{\rm PBH}\simeq3\times10^{-11}$ and $3\times10^{-9}$ for our two seed masses, respectively, and can remain relevant to a subset of high-redshift SMBHs. 

By varying the initial placement of the PBH within the same cosmological framework, we isolate the role of a large-scale environment in regulating gas inflow, BH accretion, star formation, metal enrichment, and galaxy morphology. We track the coevolution of the PBH and its host galaxy, predict their possible observational signatures and low-redshift descendants, and compare these outcomes with regimes probed by recent JWST observations and future observing campaigns. Our goal is not to associate PBHs uniquely with a single observational class, such as Little Red Dots, but to map the evolutionary pathways~\footnote{Constrained by the computational resources, we select a limited set of representative realizations to span characteristic PBH environments and initial configurations, rather than a statistical ensemble of all possible pathways.} through which PBH-seeded systems may remain exotic, temporarily appear as compact BH-dominated sources, or eventually evolve into more conventional galaxies hosting AGN.

This paper is organized as follows. In Section~\ref{sec:meth} we describe our simulation setup and analysis methodology. In Section~\ref{sec:Result} we present the structural and chemical properties of the simulated systems and compare them with observations. We discuss implications and limitations in Section~\ref{subsec:Caveat}, and offer our conclusions in Section~\ref{sec:Summary}.

\begin{table*}[htb!]
    \centering
    \caption{Key parameters and main simulation results.
    $z_{\rm ini}$ denotes the starting redshift of the simulation, while $z_{\rm final}$ corresponds to the final saved snapshot. $N_{\rm eff}$ denotes the equivalent uniform particle resolution for the initial box of size $L=2\,\mathrm{cMpc}/h$. For the full-box simulations, this corresponds to the actual particle resolution, whereas for the refined-region ('\texttt{\_2sigzoom}') simulations it represents the uniform-box resolution that would give the same particle mass as in the highest-resolution region.
$\Delta M_\bullet$ represents the total accreted mass by $z_{\rm final}$, and $\langle f_{\rm edd}\rangle \equiv \langle \dot{M}_{\bullet}/\dot{M}_{\rm edd}\rangle$ gives the time-averaged Eddington ratio between $z_{\rm ini}$ and $z_{\rm final}$. The (co-moving) softening lengths for gas and stellar particles are given by $\epsilon_{\rm gas}$, and $\epsilon_{\star}$, as well as for dark matter, $\epsilon_{\rm DM}=0.1~\rm kpc/\it h$, throughout. $M_{\rm h, PBH}$ is the mass of the halo chosen to host the PBH, at the last snapshot of the simulation. }
    \begin{tabular}{ccccccccc}
    \hline
        Run &$z_{\rm ini}$  & $z_{\rm final}$& $N_{\rm eff}$  & $\Delta M_\bullet$  &$\langle f_{\rm edd}\rangle$  & $\epsilon_{\rm gas}\small[\rm kpc/\it h]$ & $\epsilon_{\star}\small[\rm kpc/\it h]$ & $M_{\rm h, PBH}\small[M_{\odot}/h]$\\

    \hline
    
    \texttt{DM\_only}   & 3400  & 3.0 & $ 256^3$  & -& -  & - & - & $5.5 \times 10^9$ \\\texttt{DM\_only\_2sig}   & 3400  & 3.0 & $ 256^3$  & -& -  & - & - & $2.7 \times 10^{10}$ \\
    \texttt{PBH\_DMonly}   & 3400  & 1100 & $ 256^3$  & -& -   & - & - & -  \\
    \texttt{PBH\_DMonly\_dense}   & 3400  & 1100 & $ 256^3$  & -& -   & - & -  & - \\
    \texttt{PBH\_DMonly\_2sig}   & 3400  & 1100 & $ 256^3$  & -& -   & - & -  & - \\
    \texttt{PBH\_M1e7}   & 1100 & 3.8 &  $2\times 256^3$  &  $2.0\times10^6$ & 0.0038 & 0.1 & 0.01 & $8.5 \times 10^8$ \\
    \texttt{PBH\_M1e7\_dense}   & 1100 & 4.1 &  $2\times 256^3$  &  $9.0\times10^6$ & 0.011 & 0.1 & 0.01 & $8.3\times 10^9$ \\\texttt{PBH\_M1e7\_2sig}   & 1100 & 11.5 &  $2\times 256^3$  &  $4.2\times10^6$ & 0.023 & 0.1 & 0.01 & $1.3\times 10^9$ \\

        \hline
    
    \texttt{PBH\_DMonly\_M1e7\_2sigzoom}   & 3400  & 1100 & $ 256^3$  & -& -   & - & -  & - \\
    \texttt{PBH\_DMonly\_M1e5\_2sigzoom}   & 3400  & 1100 & $ 256^3$  & -& -    & - & -  & - \\
    \texttt{PBH\_DMonly\_M1e5\_2sigzoom\_2}   & 3400  & 1100 & $ 256^3$  & -& -   & - & -  & - \\
    \texttt{PBH\_M1e7\_2sigzoom}   & 1100 & 8.1 &  $2\times 256^3$  &  $8.0\times10^6$ & 0.029  & 0.1 & 0.01 & $6.0\times 10^{9}$ \\
    \texttt{PBH\_M1e5\_2sigzoom}   & 1100 & 8.7 &  $2\times 256^3$  &  $8.2\times 10^2$ & $0.00026$  & 0.1 & 0.01 & $1.1\times 10^{9}$ \\
    \texttt{PBH\_M1e5\_2sigzoom\_2}   & 1100 & 8.8 &  $2\times 256^3$  &  $1.9\times 10^3$ & $0.00094$  & 0.1 & 0.01 & $6.4\times 10^{9}$ \\

    \hline

    \hline
    \end{tabular}
    \label{Table:SimParam}
\end{table*}

\section{Methodology} \label{sec:meth}

We adopt the numerical framework developed in our previous PBH simulations~\citep{ZhangPBHGalaxy2025ApJ,ZhangDCBHPBH2025ApJ,Zhang2026PBHApJL}. All simulations are performed with the \textsc{GIZMO} code \citep{Hopkins2015MNRAS.450...53H}, which combines an updated version of the \textsc{GADGET-3} TreePM gravity solver \citep{springel2005cosmological} with the Lagrangian meshless finite-mass (MFM) hydrodynamics method using $N_{\rm ngb}=32$ neighbors. Primordial chemistry and cooling are followed with a 12-species
non-equilibrium network \citep{Bromm2002ApJ...564...23B,
Johnson2006MNRAS.366..247J}, initialized at $z=1100$ using the abundances of
\citet{Galli2013ARA&A..51..163G}. We additionally account for the fine-structure cooling of C~\textsc{II}, O~\textsc{I}, Si~\textsc{II}, and Fe~\textsc{II} \citep{Safranek2010,Jaacks2018}. We describe the new environmental initial conditions in Section~\ref{subsec:ini} and provide the adopted BH, star-formation, and feedback prescriptions in Section~\ref{subsec:sim}..

Throughout this work, the \textit{Planck18} cosmological parameters are adopted~\citep{Plank2020A&A...641A...6P}: $\Omega_{\rm m}=0.3111$, $\Omega_{\rm b}=0.04897$, $h=0.6766$, $\sigma_8=0.8102$, and $n_{\rm s}=0.9665$. The key simulation parameters are summarized in Table~\ref{Table:SimParam}.

\subsection{Initial Conditions and Environmental Setup\label{subsec:ini}}

In previous work \citep{ZhangPBHGalaxy2025ApJ, ZhangDCBHPBH2025ApJ, Zhang2026PBHApJL}, we focused on the evolution of PBHs in relatively isolated environments within $L \leq 1\,\mathrm{cMpc}/h$ boxes, where the dynamics are largely governed by the seed effect \citep{Mack2007ApJ}. However, as highlighted in \citet{ZhangPBH2024ApJ}, the evolution of PBHs over cosmic time is also expected to depend sensitively on their large-scale environment. To investigate this aspect, we enlarge the simulation volume to a comoving box of side length $L = 2\,\mathrm{cMpc}/h$ with $N=256^3$ dark matter particles generated from the \textsc{music} code~\citep{hahn2011multi}, and vary the initial placement of the PBH within this volume~\footnote{We present the physical assumption for this choice of single-seeded initial conditions in the Appendix~\ref{Appendix:A}.}.

In our fiducial setup, following \citet{Zhang2026PBHApJL}, a $M_{\bullet}\simeq 10^7\,M_{\odot}$ PBH is placed at the center of the simulation box, corresponding to a region of approximately average density~\footnote{This is confirmed by the corresponding \texttt{DM\_only} simulation without any PBH, which shows no massive halo forming near the box center.}. Here, we follow prescribed massive PBH seeds without predicting their formation probability or cosmic abundance. The system is first evolved from $z = 3400$ to $z = 1100$ in a dark-matter-only configuration, with the PBH treated as a massive particle (the \texttt{PBH\_DMonly} run), establishing the initial dark matter distribution. At $z = 1100$, baryons are introduced, and the simulation transitions to full hydrodynamics, allowing us to follow gas accretion, star formation, and feedback (the \texttt{PBH\_M1e7} run)~\footnote{Before the recombination epoch, baryons remain tightly coupled to the background radiation field through Thomson scattering, which suppresses their infall through radiation pressure and Compton drag~\citep{Carr2018MNRAS.478.3756C}.}. 

To probe overdense environments, we construct alternative realizations in which the PBH is embedded in a region that later will collapse into a massive halo. These realizations vary the seed placement while retaining the fiducial normalization, $\sigma_8=0.8102$. In a more extreme case, we increase the fluctuation normalization throughout the box by setting $\sigma_8=2.0$ and generate an additional simulation box of the same size using the \textsc{music} code \citep{Greif2011ApJ...737...75G,JeonPISN2026ApJ..1001....3J}. This enhanced-normalization experiment explores earlier structure assembly and matter supply, rather than selecting a rare overdense region within the fiducial cosmology.

We first run a dark-matter-only simulation without PBHs (\texttt{DM\_only} and \texttt{DM\_only\_2sig} for $\sigma_8=0.8102$ and 2.0) from $z = 3400$ to $z = 3$, and identify halos using the \textsc{Rockstar} halo finder \citep{behroozi2012rockstar}. We select the most massive halo at $z = 3$ ($M_{\rm h,\ PBH} \simeq 5.5 \times 10^9$ and $\simeq 2.7 \times 10^{10}\,M_\odot/h$, respectively) and trace its Lagrangian region back to the initial conditions. The center of this Lagrangian patch defines an overdense region at early times, expected to experience enhanced mass assembly and gas inflow. We then place the PBH at this location and re-run the simulation with identical parameters, yielding the \texttt{PBH\_DMonly\_dense},  \texttt{PBH\_DMonly\_2sig}, \texttt{PBH\_M1e7\_dense}, and \texttt{PBH\_M1e7\_2sig} runs. This setup enables a controlled comparison between PBHs evolving in typical and highly overdense environments, isolating the impact of large-scale structure on their growth and observable properties. The hydrodynamic simulations are run from $z
= 1100$ to $z\sim 4$ ($z\sim 12$ for the 2-sigma box due to the intensive computational cost from early stellar formation and feedback), as limited by the box size and computational resources, corresponding to the redshift where the number density of observed LRDs has decreased~\citep[see e.g.,][]{Ma2026LRDApJ}, and also at a redshift similar to one of the recently discovered overly massive SMBHs, the Cliff~\citep[confirmed at $z\simeq 3.5$, see][]{Cliff2025A&A, Cliff2026MNRAS}. 
We note that as a consequence of our choice of initial conditions, the PBH system will start life as an overmassive BH embedded in a star forming halo.
To partially mitigate the computational cost of the full \texttt{PBH\_M1e7\_2sig} run, we also construct a refined-region realization centered on the same overdense Lagrangian patch. This setup is generated using the same zoom-in initial-condition machinery in \textsc{music}, but is used here primarily to focus the computational resolution on the PBH-hosting overdense region rather than to perform a full multi-scale zoom-in study. Specifically, we use the \texttt{DM\_only\_2sig} run to identify the center and spatial extent of the Lagrangian region associated with one of the most massive halos at $z=3$. Instead of evolving the full $L=2\,{\rm cMpc}/h$ volume at uniform resolution, we generate refined initial conditions at $z=3400$ for the dark-matter-only stage and at $z=1100$ for the hydrodynamical stage. The refined region has a side length of $L_{\rm ref}\sim750\,{\rm ckpc}/h$, approximately corresponding to the extent of the Lagrangian patch, in which the resolution is identical to that of the full \texttt{PBH\_M1e7\_2sig} run, while outside the refined region, the mass resolution is reduced by a factor of $4^3$, and there are no gas particles in the hydrodynamical stage~\footnote{This particular choice of the simulation extent also aims at reducing the influence of external gas inflow to the PBH host halo. In all ``zoom'' runs, the PBH host remains well inside the gas-populated region, with the minimum distance from its center to the outer gas boundary exceeding $8R_{\rm vir}$ down to $z\simeq8$ where the simulation ends.}. 
The corresponding dark-matter-only run, \texttt{PBH\_DMonly\_M1e7\_2sigzoom}, is evolved from $z=3400$ to $z=1100$ to establish the initial dark-matter configuration for the hydrodynamical run, \texttt{PBH\_M1e7\_2sigzoom}, following the same procedure as above. In addition, we test the effect of varying the initial PBH seed mass. Using the same Lagrangian-region selection and refined-region procedure, we place a lighter PBH seed of mass $10^5\,M_\odot$ in the identified overdense region of the \texttt{DM\_only\_2sig} box. This gives rise to four additional simulations, \texttt{PBH\_DMonly\_M1e5\_2sigzoom}, \texttt{PBH\_DMonly\_M1e5\_2sigzoom\_2}, \texttt{PBH\_M1e5\_2sigzoom}, and \texttt{PBH\_M1e5\_2sigzoom\_2}. Here the label ``zoom'' is retained for bookkeeping, although these runs should be interpreted as refined-region realizations of the overdense PBH environment. The runs with the suffix ``\texttt{\_2}'' differ by initializing the $10^5\,M_\odot$ PBH at the center of the Lagrangian region of a $1.8\times10^9\,M_{\odot}/h$ halo at $z=7$, which is a progenitor of the main halo identified at $z=3$. As a result, the PBH host halo merges into the final system at an earlier stage than in the corresponding runs without the ``\texttt{\_2}'' suffix.

In all hydrodynamic runs listed in Table~\ref{Table:SimParam} (\texttt{PBH\_M1e7}, \texttt{PBH\_M1e7\_dense}, \texttt{PBH\_M1e7\_2sig}, \texttt{PBH\_M1e7\_2sigzoom}, \texttt{PBH\_M1e5\_2sigzoom}, and \texttt{PBH\_M1e5\_2sigzoom\_2}), the dark matter and gas particle masses are $5.1\times10^4\,M_\odot$ and $9.6\times10^3\,M_\odot$, respectively. This resolution pertains to the refined region in the respective runs.

\subsection{Star Formation and Feedback Physics\label{subsec:sim}}

The BH accretion and thermal/radiative feedback subgrid models are implemented using the numerical prescriptions from our previous work~\citep{ZhangPBHGalaxy2025ApJ, Zhang2026PBHApJL}. The growth of the central PBH is self-consistently tracked and computed using Bondi--Hoyle--Lyttleton accretion, and feedback is implemented by depositing a fraction of the accretion energy into the surrounding gas. We adopt a fiducial coupling efficiency of $\epsilon_r = \Delta E_{\bullet, \rm inj} / (L_{\bullet, \rm acc} \Delta t) = 0.005$, as a value capable of reproducing some of the overly massive and pristine BHs observed by JWST~\citep{Zhang2026PBHApJL}~\footnote{The impact of varying this parameter has been explored in previous work; see~\citet{ZhangPBHGalaxy2025ApJ,Zhang2026PBHApJL} for details.  Further discussions for the effect of this parameter in overdense/enhanced-normalization cosmic environments are included in Appendix~\ref{Appendix:B}.}. Here, $L_{\bullet,\rm acc}$ is the accretion luminosity, and $\Delta E_{\bullet,\rm inj}$ is the energy injected into the ambient gas over the timestep $\Delta t$. 
A drag force was also applied to BH particles during accretion to conserve momentum according to ~\cite{springel2005modelling}. This choice mitigates numerical kicks associated with very strong accretion episodes and helps prevent artificial BH wandering from the galaxy center at lower redshift.

As the gas will aggregate around the PBH-seeded halo, star formation occurs when gas becomes Jeans unstable and survives both BH accretion and feedback-driven dispersal long enough to undergo free-fall collapse. The stellar population is assigned according to the gas metallicity at formation, with Pop~III and Pop~II stars separated by a critical metallicity threshold $Z_{\rm th}=10^{-4}\,Z_\odot$ \citep[e.g.,][]{BFCL2001}, where $Z_\odot=0.0134$. 
Each star-forming gas particle spawns 16 stellar particles, corresponding to an individual mass of $\simeq600\,M_\odot$.  For Pop~III stars, each stellar particle represents a Pop~III cluster with a characteristic mass consistent with those expected from H$_2$-cooling~\citep[e.g.,][]{Stacy2013PopIII,Hirano2017MNRAS.470..898H,Liu2021, Liu2024_Mass,Gurian2026}. Pop~III stellar masses are drawn on-the-fly from a modified Larson IMF, $dN / dM \propto M^{-\alpha} \exp(-M_{\rm cut}^2 / M^2)$, over a mass range of $1-150 \,M_{\odot}$, adopting $\alpha = 0.17$ and $M_{\rm cut}^2 = 20 \,M_{\odot}^2 $~\citep{Jaacks2018}. The mass distribution for Pop~II stars follows a Chabrier IMF, with a mass range of $0.08 - 100\,M_{\odot}$~\citep{Jaacks2019}.

The stellar feedback is composed of ionization heating, Lyman-Werner (LW) radiation that dissociates $\rm H_2$, supernova (SN) feedback that includes thermal energy injection, and metal enrichment~\citep{Jaacks2018, Jaacks2019, Liu2020}. Specifically, the LW background is computed from the sum of a global star formation rate density contribution and local stellar sources, assuming an optically thick regime with self-shielding included~\citep{Liu2020}~\footnote{We do not include the LW contribution from BH accretion in this work; its possible impact has been discussed in \citet{ZhangPBHGalaxy2025ApJ,ZhangDCBHPBH2025ApJ}}. We model intergalactic medium (IGM) photo-ionization  using a spatially uniform UV background based on the calculation of \citet{FaucherUVB2009}, with a characteristic self-shielding scale of $\sim1\,{\rm kpc}$.  After characteristic lifetimes of $\sim3$ Myr for Pop~III particles and $\sim20$ Myr for Pop~II particles, SN feedback is applied following the prescription of \citet{Jaacks2018,Liu2020}.

Since individual blast waves are unresolved, we adopt a subgrid SN legacy model in which the 
metal yields are deposited into gas particles within a characteristic final shell radius of $\sim650\,{\rm pc}$. The adopted metal yields are $\sim39\,M_\odot$ for Pop~III particles and $\sim10\,M_\odot$ for Pop~II ones. 
For Pop~III particles, we also impose energy injection of $\sim7\times10^{51}\,{\rm erg}$ and instantaneous ionization in the SN bubble \citep{Liu2020MNRAS.497.2839L}. 
In addition, we include SN-driven winds following \citet{Springel2003SPH} for both Pop~III and Pop~II: gas particles associated with star formation are stochastically launched with mass-loading factor $\eta_{\rm w,SF}=2$ and kick velocity $v_{\rm w,SF}\simeq170\,{\rm km\,s^{-1}}$. Wind particles recouple to the ISM after $t_{\rm w}=0.1H^{-1}(t)$, or once their density falls below $n_{\rm w}=10\,{\rm cm^{-3}}$. These prescriptions allow unresolved SN feedback to regulate star formation, metal enrichment, and gas removal in the simulated halos. For a more detailed description of the stellar feedback recipe, one can refer to~\citet{Liu2020}.

\section{Results and Discussions} \label{sec:Result}

Based on the initial conditions and numerical setup described above, we now present the simulation results and discuss their physical implications. To highlight the environmental dependence of PBH-seeded evolution, we organize this section around four connected aspects. We first examine how the large-scale environment drives distinct BH--host evolutionary pathways in Section~\ref{subsec:PBHBranches}. Building on these trends, we identify the main evolutionary phases and possible transitions toward normal galaxy--AGN configurations in Section~\ref{subsec:normal}. We then connect these regimes to stellar growth and metal enrichment in Section~\ref{subsec:SFH&Metal}  before translating the resulting evolutionary stages into observable signatures for current and future surveys in Section~\ref{subsec:obs}.


\subsection{Environment-dependent Evolution\label{subsec:PBHBranches}}


\begin{figure}[htb!]
    \centering
    \includegraphics[width=\linewidth]{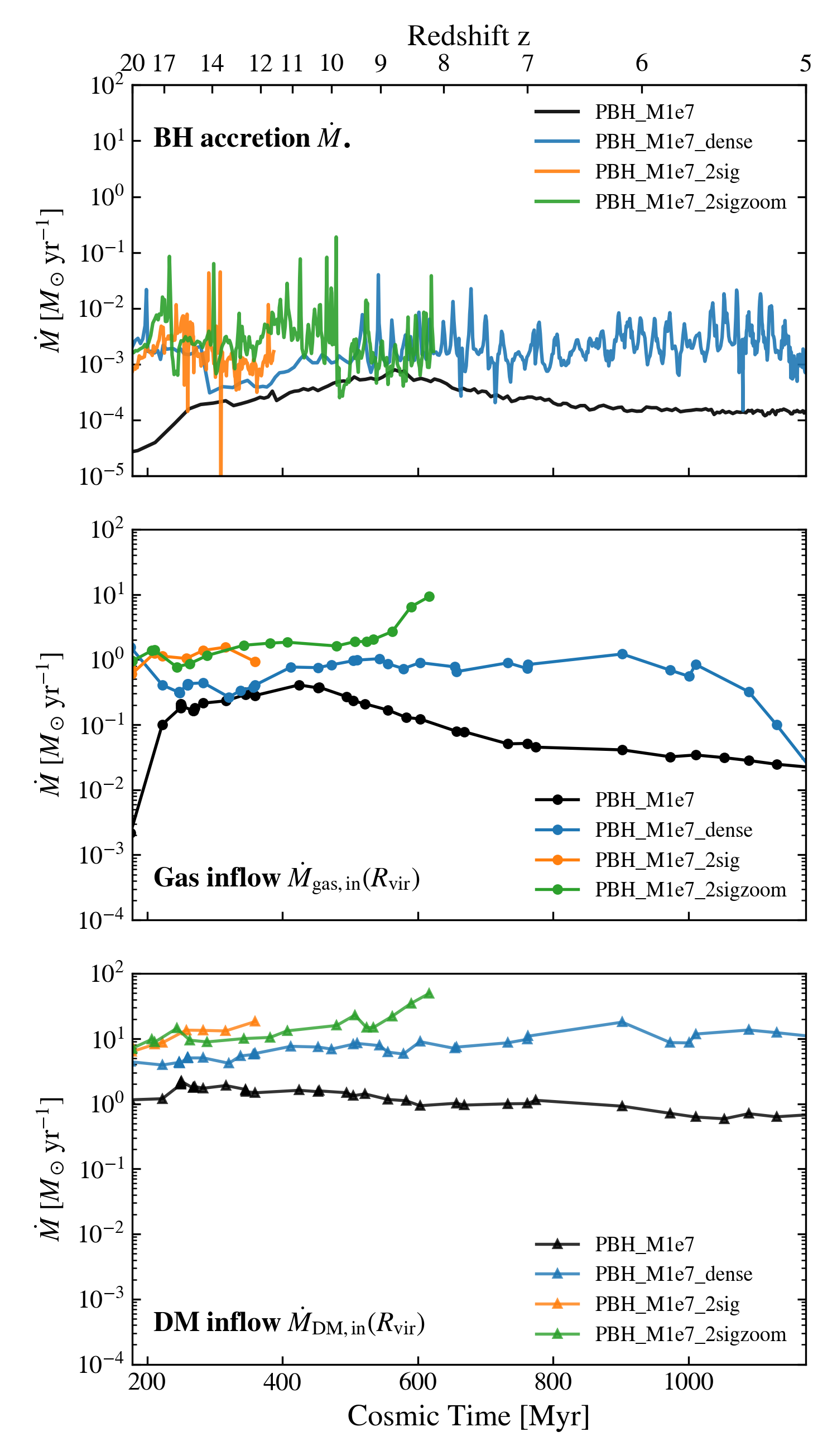}
    \caption{Comparison between BH accretion and halo-scale mass inflow for the $10^7\,M_\odot$ PBH runs. The top panel shows the BH accretion rate, $\dot{M}_\bullet$, while the middle and bottom panels show the gas and dark-matter inflow rates measured at the virial radius, $\dot{M}_{\rm gas,in}(R_{\rm vir})$ and $\dot{M}_{\rm DM,in}(R_{\rm vir})$. The bottom axis gives cosmic time, and the top axis gives the corresponding redshift. The overdense/enhanced-normalization runs show larger halo-scale inflow and more bursty BH accretion than the isolated case, illustrating the connection between structure formation and PBH fueling.}
    \label{fig:accrethist}
\end{figure}

\begin{figure*}[htb!]
    \centering
    \includegraphics[width=\linewidth]{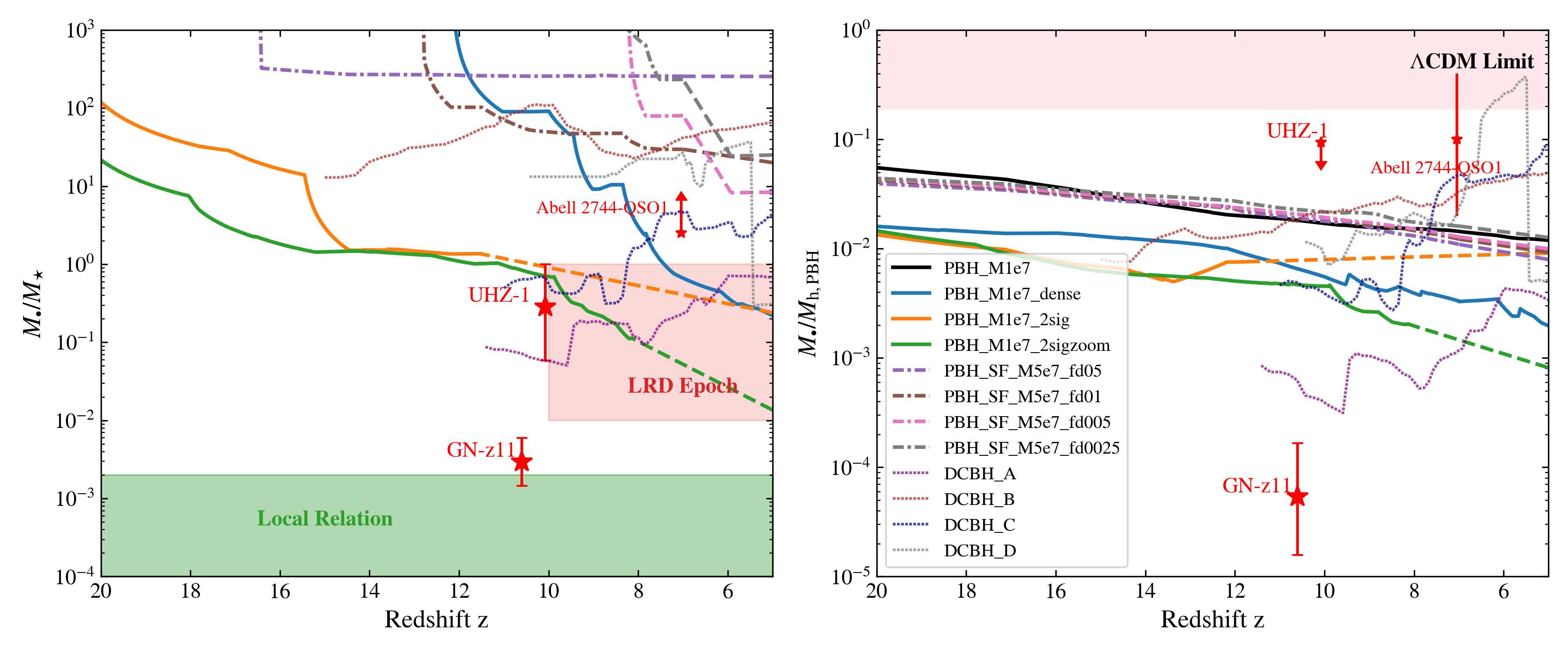}
    \caption{Evolution of the black-hole-to-stellar mass ratio, $M_{\bullet}/M_\star$ (left), and black-hole-to-halo mass ratio, $M_{\bullet}/M_{\rm h, PBH}$ (right), as a function of redshift.
Solid curves show the PBH-seeded simulations in different environments presented in this work, while dash-dotted curves show the ``\texttt{\_M5e7\_}'' runs with different feedback prescriptions in a cosmic-average environment from \citet{Zhang2026PBHApJL}; dashed extensions indicate linear extrapolations of the mass ratio down to $z=5$ if the stopping redshift $z_{\rm final}>5$, representing the evolution trend indicated by those runs. Dotted curves show representative DCBH models from \citet{JeonSMBH2025ApJ...979..127J}. 
 Observed estimates for GN-z11, UHZ-1, and Abell~2744-QSO1 are marked by red stars \citep[values adopted from][]{Natarajan:2023UHZ1,Scholtz2024A&A, Maiolino2024NaturBH,QSO1Direct2026Natur,KashlinskyPBH2026MNRAS}. In the left panel, the shaded regions indicate the approximate parameter space associated with LRD-like systems~\citep[e.g.,][]{MaiolinoBH2024A&A, KocevskiLRD2025ApJ...986..126K} and the typical BH--stellar mass relation from local observations~\citep{Kormendy2013GalaxySMBH}. In the right panel, the shaded band marks the approximate $\Lambda$CDM limit for any baryonic component, $M_{\bullet}/M_{\rm h, PBH} \lesssim \Omega_b / ( \Omega_{m} - \Omega_{b})$. The isolated PBH runs remain highly overmassive relative to their stellar components because star formation is inefficient, whereas PBHs in overdense environments undergo more rapid stellar assembly and evolve toward lower $M_{\bullet}/M_\star$. By $z\lesssim10$, some PBH systems born in overdense/enhanced-normalization regions reach less extreme mass ratios compared to high-redshift BH-dominated sources, while continuing structure growth can drive them toward more conventional galaxy--AGN configurations.}
    \label{fig:mbh-mstar}
\end{figure*}

Our previous simulations approached PBH evolution from two complementary limits. 
In the full hydrodynamical runs, the PBH was placed near the center of the simulation volume, allowing us to isolate the local seed effect on gas accretion, star formation, and feedback in a region close to the cosmic mean density~\citep{ZhangPBHGalaxy2025ApJ, ZhangDCBHPBH2025ApJ, Zhang2026PBHApJL}. In separate dark-matter-only calculations, PBHs were instead placed according to the density fluctuation field, showing that they can become embedded in larger-scale structures or in halos whose masses exceed the characteristic PBH-induced halo scale~\citep{ZhangPBH2024ApJ}, given by $\tilde{M}_{\rm h,PBH} \propto (z_{\rm eq} /z) M_\bullet $, according to the analytic estimate of \citet{Mack2007ApJ}. The present simulations connect these two approaches by following the baryonic evolution of PBHs placed in different initial cosmological environments.

Figure~\ref{fig:accrethist} compares the BH accretion history with the halo-scale gas and dark-matter inflow rates for the $10^7\,M_\odot$ PBH runs. In the isolated \texttt{PBH\_M1e7} case, the BH accretion rate (top panel) first rises to $\dot{M}_\bullet\sim  10^{-3}\,M_\odot\,{\rm yr}^{-1}$ around $z\sim8$ -- $9$ ($f_{\rm Edd}\sim10^{-2}$), and then gradually declines toward $\dot{M}_\bullet\sim10^{-4}\,M_\odot\,{\rm yr}^{-1}$ at later times ($f_{\rm Edd}\sim10^{-3}$). This decline follows the reduction of the gas inflow rate (middle panel) through the virial radius, which decreases from $\dot{M}_{\rm gas,in}(R_{\rm vir})\sim 3\times10^{-1}\,M_\odot\,{\rm yr}^{-1}$ at early times to $\lesssim10^{-1}\,M_\odot\,{\rm yr}^{-1}$ toward the end of the run. The corresponding dark-matter inflow rate (bottom panel) is relatively modest, remaining at the level of $\dot{M}_{\rm DM,in}(R_{\rm vir})\sim1\,M_\odot\,{\rm yr}^{-1}$ or below, indicating slow halo growth in this environment.

By contrast, the overdense runs show systematically larger halo-scale inflow and more bursty BH fueling. For \texttt{PBH\_M1e7\_dense}, the gas inflow rate remains at $\dot{M}_{\rm gas,in}(R_{\rm vir})\sim0.3$ -- $1\,M_\odot\,{\rm yr}^{-1}$ over an extended period, while the dark-matter inflow rate is typically $\dot{M}_{\rm DM,in}\sim4-10\,M_\odot\,{\rm yr}^{-1}$. The BH accretion rate correspondingly fluctuates around $\dot{M}_\bullet\sim10^{-3}$ -- ${\rm few}\times10^{-2}\,M_\odot\,{\rm yr}^{-1}$ ($f_{\rm Edd}\sim10^{-2}$ -- $10^{-1}$), with short bursts reaching higher values. The \texttt{PBH\_M1e7\_2sig} and \texttt{PBH\_M1e7\_2sigzoom} runs, although followed only to higher redshift, exhibit the same behavior at earlier cosmic times than the \texttt{PBH\_M1e7\_dense} run, with elevated gas and dark-matter inflow associated with rapid halo assembly. These results show that the different BH accretion histories are connected to the larger-scale supply of matter: overdense environments continue to deliver gas to the PBH-hosting halo, whereas the isolated case is more easily starved once the initial gas reservoir is depleted or affected by feedback.

The same environmental separation appears in the mass-ratio evolution shown in Figure~\ref{fig:mbh-mstar}. We track both the BH-to-stellar mass ratio, $M_\bullet/M_\star$, and the BH-to-halo mass ratio, $M_\bullet/M_{\rm h,PBH}$, as functions of redshift. For reference, the green shaded region indicates the approximate range for the local BH-to-stellar mass relation at $M_\bullet/M_\star\lesssim 0.002$~\citep{Kormendy2013GalaxySMBH}, while the red shaded region marks the broad range of ratios at $\sim 0.01 -1$ associated with compact, BH-dominated systems~\citep[e.g.,][]{MaiolinoBH2024A&A,KocevskiLRD2025ApJ...986..126K}. Representative measurements for GN-z11, UHZ-1, and Abell~2744-QSO1 are also shown for comparison \citep[values adopted from][]{Natarajan:2023UHZ1,Scholtz2024A&A, Maiolino2024NaturBH,QSO1Direct2026Natur,KashlinskyPBH2026MNRAS}.

In the isolated \texttt{PBH\_M1e7} run, star formation remains inefficient throughout the simulation due to strong accretion heating and inefficient gas inflow, similar to the behavior of the ``\texttt{PBH\_SF\_M5e7}'' models previously used to study Abell~2744-QSO1 in \citet{Zhang2026PBHApJL}. The system maintains an extreme mass ratio, $M_\bullet/M_\star\geq 4\times 10^4$, from the onset of galaxy formation down to $z\sim4$, where the ratio reaches its minimum value. It therefore remains substantially more BH dominated than the observed high-redshift systems. Its $M_\bullet/M_\star$ track lies above the plotting range of Figure~\ref{fig:mbh-mstar} at all times and is therefore not visible in the left panel. The absence of sustained stellar assembly prevents this branch from evolving toward regimes with lower mass ratios.

In the overdense \texttt{PBH\_M1e7\_dense} run, the PBH is supplied by a denser environment and accretes more efficiently, but the stellar and halo components grow substantially faster. Star formation occurs both within the PBH host halo and in nearby minihalos that subsequently merge with the central system. As a result, $M_\bullet/M_\star$ decreases by several orders of magnitude, evolving from an initially extreme PBH-dominated state at $z\gtrsim9$ to $M_\bullet/M_\star\sim0.1$ by $z\sim5$. The evolution of $M_\bullet/M_{\rm h,PBH}$ follows the same qualitative trend: overdense environments allow the host halo to grow quickly enough that the PBH becomes a progressively smaller fraction of the total halo mass.

In the enhanced-normalization runs ($\sigma_8=2.0$), \texttt{PBH\_M1e7\_2sig} and \texttt{PBH\_M1e7\_2sigzoom}, exhibit an even earlier onset of star formation, beginning at $z\sim30$. Because the surrounding structure collapses earlier and supplies the PBH host more efficiently, the stellar component grows rapidly and $M_\bullet/M_\star$ reaches the approximate range $0.1$--$1$ by $z\sim11$--$8$. These tracks therefore enter the regime occupied by compact high-redshift BH hosts earlier than the less overdense branch and continued stellar assembly subsequently drives them toward lower mass ratios. These track pass through the range occupied by UHZ-1 and other compact high-redshift BH-dominated systems, although they remain above the local relation by the end of the simulation.

For comparison, a version of DCBH models explored by \citep{JeonSMBH2025ApJ...979..127J,JeonLRD2026ApJ...998..148J} exhibit qualitatively different evolutionary behavior. In these models, star formation in a neighboring halo first establishes the radiative conditions required for direct collapse \citep[e.g.,][]{Visbal2014,Baggen2026}. The DCBH therefore forms only after baryonic structure formation has begun, in contrast to the PBH seeds, which are present before the formation of their host galaxies. Across the snapshots in which the DCBHs are identified, the systems span $M_\bullet/M_\star\sim0.01$--$10$ and subsequently show comparatively fluctuating evolution without a common monotonic trend~\footnote{The first plotted DCBH point corresponds to the first saved snapshot in which the BH is identified, rather than its exact formation time.}. More pronounced decreases occur when the DCBH host merges with a neighboring star-forming galaxy, whose stellar mass rapidly lowers $M_\bullet/M_\star$.


\subsection{Road to Normalcy} \label{subsec:normal}

\begin{figure}[htb!]
    \centering
    \includegraphics[width=\linewidth]{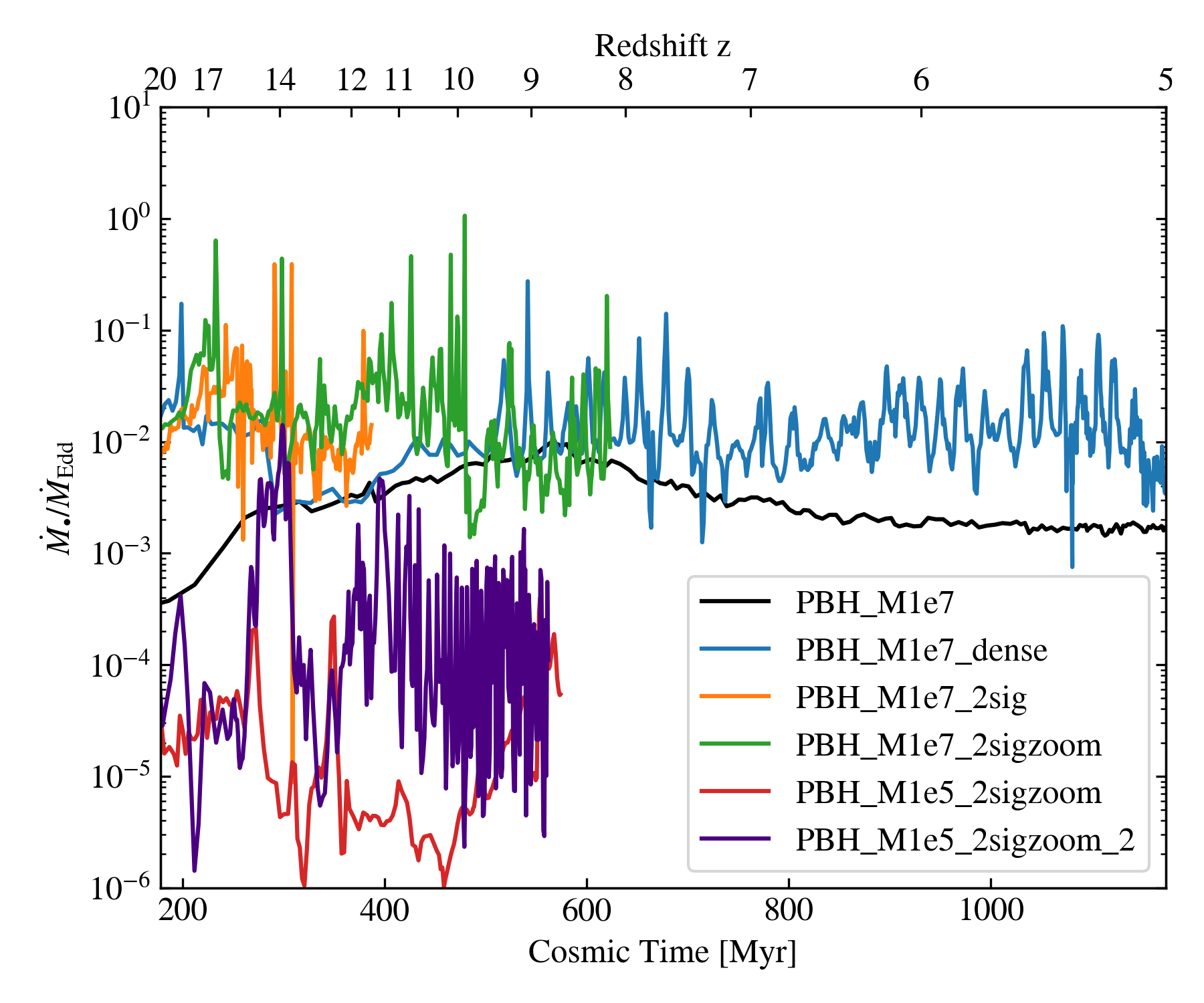}
\caption{
Eddington-normalized BH accretion histories, extending Figure~\ref{fig:accrethist} to the refined-region realizations and the $10^5\,M_\odot$ seed runs. The $10^7\,M_\odot$ PBHs generally accrete at $\dot{M}_\bullet/\dot{M}_{\rm Edd}\sim10^{-3}$--$10^{-2}$ with intermittent bursts, whereas the lighter seeds remain at lower Eddington ratios. The centrally placed \texttt{PBH\_M1e5\_2sigzoom\_2} seed nevertheless experiences stronger and more frequent fueling than the corresponding offset realization, illustrating the additional dependence on seed placement and halo assembly history.
}
    \label{fig:mdotzoom}
\end{figure}

\begin{figure*}[htb!]
    \centering
    \includegraphics[width=\linewidth]{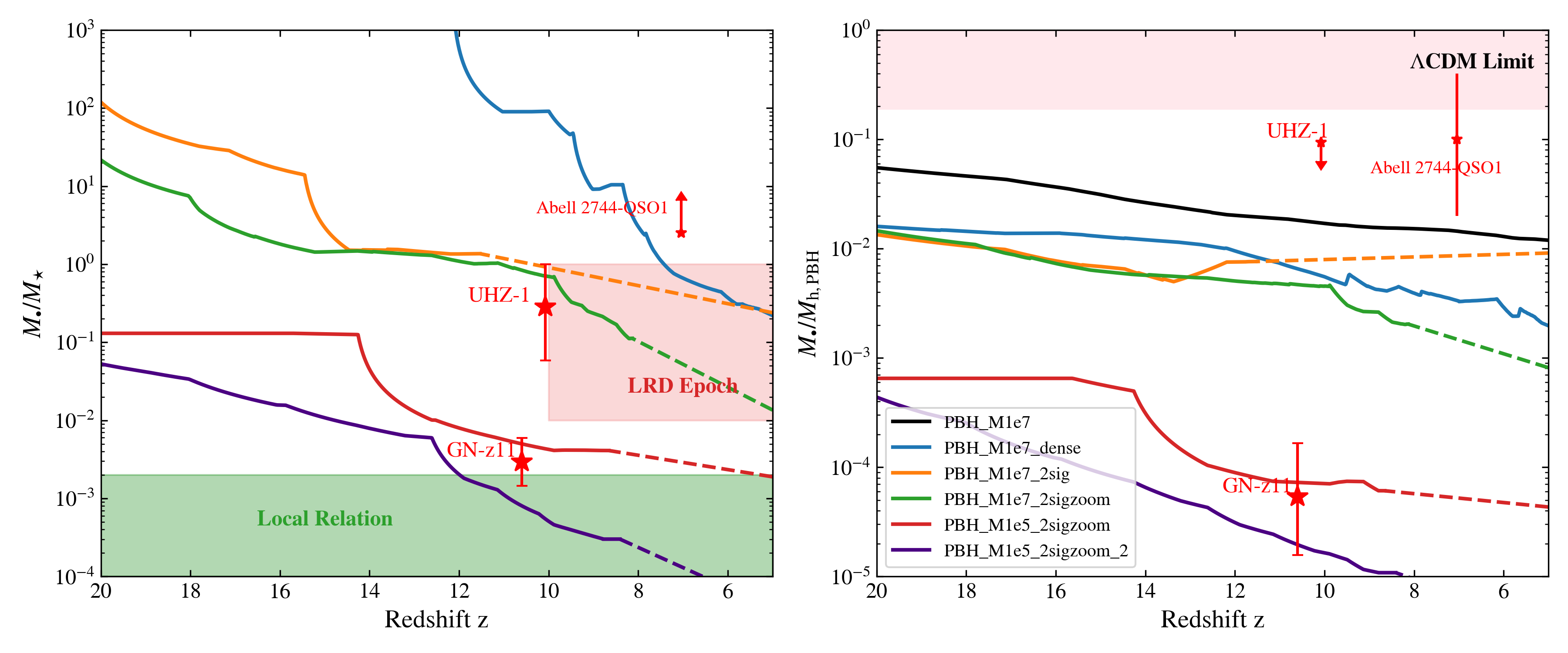}
\caption{
Evolution of the BH-to-stellar mass ratio, $M_{\bullet}/M_\star$ (left), and the BH-to-host-halo mass ratio, $M_{\bullet}/M_{\rm h,PBH}$ (right), for the full-volume and refined-region simulations spanning different PBH seed masses and initial placements within overdense/enhanced-normalization environments. The observational reference regions and individual high-redshift systems are the same as in Figure~\ref{fig:mbh-mstar}. The $10^5\,M_\odot$ seed realizations evolve toward lower mass ratios than the corresponding $10^7\,M_\odot$ cases, as stellar and halo assembly more rapidly outpace BH growth. Similar to Figure~\ref{fig:mbh-mstar}, dashed lines represent the extrapolated trend from individual runs. The difference between the two $10^5\,M_\odot$ realizations further illustrates the sensitivity of the evolutionary track to the initial PBH location and its subsequent incorporation into the assembling main halo.}
    \label{fig:mbh-mstarzoom}
\end{figure*}

The results of Section~\ref{subsec:PBHBranches}  show that the cosmological environment regulates the gas supply and co-evolution of PBH-hosting systems. We now examine how the initial PBH mass and its placement within the assembling overdense region modify this evolution. Here, ``normalcy'' is defined in an evolutionary, trend-driven sense. It does not require exact agreement with the local BH--galaxy scaling relations. Instead, it refers to a progressive reduction in the relative dominance of the initial seed, such that stellar and halo growth outpace BH growth and drive both $M_\bullet/M_\star$ and $M_\bullet/M_{\rm h,PBH}$ toward the ranges occupied by more conventional high-redshift galaxy--AGN systems.

Figure~\ref{fig:mdotzoom} shows that different seed realizations experience distinct fueling histories even within broadly similar overdense environments. The $10^7\,M_\odot$ PBHs typically accrete at $f_{\rm Edd}\sim10^{-3}$--$10^{-2}$ during the main assembly phase, with intermittent bursts reaching $f_{\rm Edd}\sim0.1$ or higher in the overdense/enhanced-normalization runs. By contrast, the offset $10^5\,M_\odot$ seed in \texttt{PBH\_M1e5\_2sigzoom} remains weakly fueled for most of its evolution, typically at $f_{\rm Edd}\sim10^{-6}$--$10^{-4}$. The corresponding ``\texttt{\_2}'' realization experiences more frequent episodes at $f_{\rm Edd}\sim10^{-4}$--$10^{-3}$, with occasional excursions to $\sim10^{-2}$. This difference reflects the initialization of the ``\texttt{\_2}'' seed near the center of a halo that is subsequently incorporated into the assembling main system, allowing it to remain more closely coupled to the available gas supply. BH growth remains modest in many cases due to accretion feedback and lack of dense gas, with the cumulative mass increase of $\Delta M_\bullet(z_{\rm final})/M_\bullet(z_{\rm ini})\lesssim 1$ by the end of the simulations, where
$\Delta M_\bullet (z)=M_\bullet (z)-M_{\bullet}(z_{\rm ini})$. In one case where BH feedback is relatively inefficient and embedded in the center of the overdense region, $\Delta M_\bullet(z\simeq 9.5)/M_\bullet(z_{\rm ini})\simeq 4$ as demonstrated in Appendix~\ref{Appendix:B}.

The consequences for the BH-to-host mass ratios are shown in Figure~\ref{fig:mbh-mstarzoom}. As discussed above, the overdense $10^7\,M_\odot$ runs evolve away from their initially extreme, BH-dominated configurations as stellar and halo assembly proceed, although they generally remain overmassive relative to their hosts throughout the simulated interval, comparable to the observed LRD systems~\citep[e.g.,][]{MaiolinoBH2024A&A}. The $10^5\,M_\odot$ realizations follow different pathways. Their $M_\bullet/M_\star$ ratios begin at $\sim0.05$--$0.1$ at $z\sim20$ and rapidly decline toward host-dominated configurations. By $z\sim10$, \texttt{PBH\_M1e5\_2sigzoom} reaches $M_\bullet/M_\star\sim \mathcal{O}(10^{-3})$, while the more rapidly assembled ``\texttt{\_2}'' realization reaches values of $\mathcal{O}(10^{-4})$. Their BH-to-halo mass ratios similarly decline from $\mathcal{O}(10^{-4})$ to $\mathcal{O}(10^{-5})$. Although the centrally placed ``\texttt{\_2}'' seed is fueled more efficiently than the offset realization, its stellar and halo components grow even faster, producing a lower BH-to-host mass ratio comparable to that inferred for GN-z11~\citep{Bunker2023A&AGNz11,Scholtz2024A&A, Maiolino2024NaturBH}. The lighter-seed tracks therefore approach, and in the ``\texttt{\_2}'' case fall into, the local reference range shown in Figure~\ref{fig:mbh-mstarzoom}. This does not imply that these systems have already converged onto the $z=0$ scaling relation, but demonstrates that a PBH-seeded system can lose the extreme BH-to-host ratios associated with massive early seeds and appear comparatively conventional in these bulk properties by $z\sim10$--8. In contrast, the more massive PBH seeds remain recognizably overmassive for a longer period within similarly overdense environments.

Taken together, Figures~\ref{fig:mbh-mstar} and \ref{fig:mbh-mstarzoom} show that PBH-seeded systems do not follow a unique evolutionary track. The explored parameter space encompasses persistently BH-dominated systems, faint and weakly fueled seeds, and objects that rapidly develop comparatively conventional BH-to-host mass ratios. As host assembly proceeds, these bulk properties retain progressively less information about the initial PBH seed. In particular, a system that reaches the mass-ratio regime occupied by more conventional high-redshift galaxy--AGN systems may no longer retain its initial seeding conditions. The clearest imprint of the seeding conditions is therefore expected during the earlier BH-dominated or weakly coupled phases (i.e. at $z\gtrsim 10$), before subsequent stellar and halo growth drives the system toward ``normalcy''. In later stages, a single measurement of $M_\bullet/M_\star$ or $M_\bullet/M_{\rm h,PBH}$ is unlikely to recover the initial conditions without additional information on the host structure, environment, and evolutionary state.

\subsection{Star Formation and Metal Enrichment Histories\label{subsec:SFH&Metal}}


\begin{figure}[htb!]
    \centering
    \includegraphics[width=\linewidth]{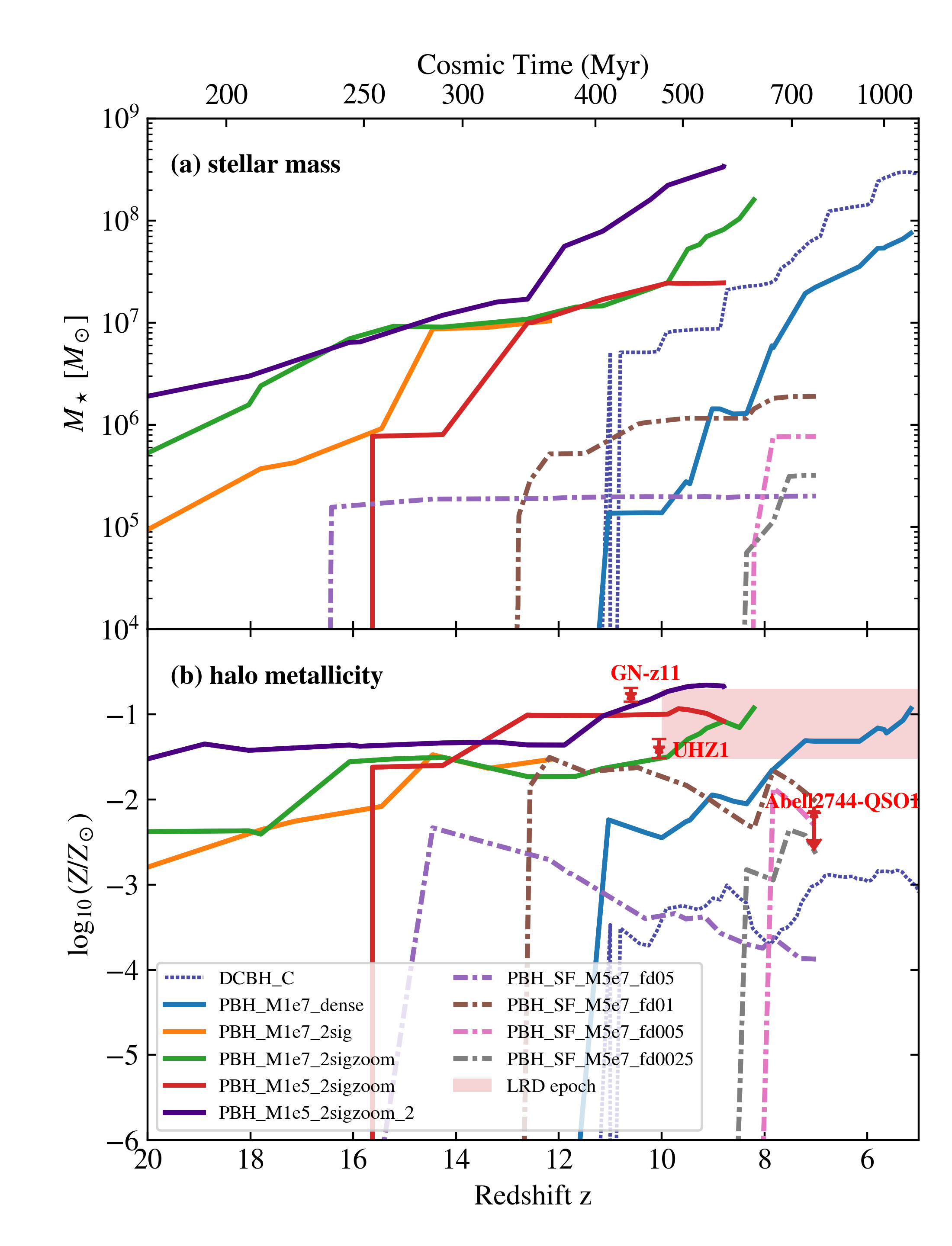}
\caption{
Evolution of stellar mass and gas metallicity of the simulated BH-hosting systems. 
The upper panel shows the total stellar mass within the host halo, while the lower panel shows the corresponding mass-weighted gas metallicity within the halo. 
The same colors and line styles denote the same simulations in both panels, and the upper axis gives the corresponding cosmic time, following Figure~\ref{fig:mbh-mstar}. Here, we further include the two ``\texttt{\_M1e5\_}'' runs with lighter PBH seeds of $M_{\bullet}=10^5\ M_\odot$. The \texttt{PBH\_M1e7} run is not shown due to its extremely low stellar mass. 
Rapid stellar-mass growth in the overdense/enhanced-normalization PBH systems in this work (solid) is accompanied by sustained chemical enrichment, whereas the BH feedback-dominated, average-density realizations from \citet{Zhang2026PBHApJL} (dashed-dotted, ``\texttt{\_M5e7\_}'' runs) undergo only brief stellar-growth episodes, followed by metal loss and dilution. 
The shaded region marks the approximate metallicity range inferred for $z>5$ LRDs, $Z\sim0.03$--$0.2\,Z_\odot$~\citep{NikopoulosLRDmetal2026arXiv}.
Representative estimates for GN-z11 and UHZ-1 are shown as stars with errorbars, while the downward arrow denotes the metallicity upper limit for Abell~2744-QSO1
\citep{Scholtz2024A&A,Natarajan:2023UHZ1,QSO1PristineMaiolino2025}.
}
    \label{fig:stellar-metallicity}
\end{figure}

The mass-ratio evolution above shows how different PBH systems move through the BH--host parameter space. We next examine the stellar growth and chemical enrichment responsible for these trajectories. Figure~\ref{fig:stellar-metallicity} compares the evolution of the stellar mass within the host halo with its mass-weighted gas metallicity. Rapid increases in $M_\star$ trace episodes of active star formation and merger-driven stellar assembly, whereas extended plateaus indicate periods of weak or negligible stellar growth. Comparing the two panels therefore illustrates how metal production, retention, and dilution respond to the different star-formation histories.

In the relatively isolated and feedback-dominated PBH systems, stellar growth occurs through only one or a few short episodes, consistent with our previous results~\citep{ZhangPBHGalaxy2025ApJ,Zhang2026PBHApJL}. The ``\texttt{PBH\_SF\_M5e7}'' models generally remain at $M_\star\sim10^5$--$10^6\,M_\odot$ for extended periods following their initial star-forming episodes. The first SNe enrich the halo gas to $Z\sim10^{-2.5}$--$10^{-1.5}\,Z_\odot$, while the central region within approximately $150\,{\rm pc}$ of the BH can temporarily reach $Z\sim0.1\,Z_\odot$. Once stellar growth stalls, however, this enrichment is not sustained. Metal-loaded outflows remove part of the newly produced metals, while subsequent halo growth is dominated by relatively pristine intergalactic inflow. The mass-weighted halo metallicity consequently decreases by approximately one to two orders of magnitude, with several tracks reaching $Z\sim10^{-4}$--$10^{-2}\,Z_\odot$ by $z\sim7$. The different tracks further illustrate the sensitivity of metal retention to the adopted BH feedback efficiency~\footnote{For the ``\texttt{PBH\_SF\_M5e7}'' runs, the digits following ``\texttt{fd}'' denote the adopted feedback efficiency; for example, ``\texttt{fd05}'' corresponds to $\epsilon_r=0.05$. As discussed in our previous work, the resulting stellar mass does not vary monotonically with feedback efficiency.}.

The overdense systems instead show sustained stellar assembly accompanied by progressively increasing metallicity. In the \texttt{PBH\_M1e7\_2sigzoom} runs, star formation begins at $z\sim28$, and the stellar mass grows from $M_\star\sim10^5$--$10^6\,M_\odot$ at $z\sim20$ to $10^7$--$10^8\,M_\odot$ by $z\sim9$ -- 8. Over the same period, the mass-weighted halo metallicity increases from $Z\sim10^{-3}$--$10^{-2}\,Z_\odot$ to $Z\sim0.1\,Z_\odot$. The \texttt{PBH\_M1e7\_dense} run follows a similar pathway at later times: its stellar mass remains below $10^6\,M_\odot$ until $z\sim10$, but subsequently grows to approximately $10^8\,M_\odot$ by $z\sim5$, while the halo metallicity rises from $Z\sim0.006\,Z_\odot$ to approximately $0.1\,Z_\odot$. Although temporary declines occur during individual inflow or feedback episodes, the overall enrichment trend remains positive. In these environments, inflowing gas is not exclusively pristine, as mergers and accretion from neighboring star-forming progenitors also supply previously enriched material. Continued in-situ star formation and enriched inflow therefore compensate for dilution and feedback-driven metal loss.

The lighter-seed simulations show comparable or even stronger stellar growth among the overdense PBH models compared with the heavier seeded runs. In \texttt{PBH\_M1e5\_2sigzoom\_2}, $M_\star$ increases from approximately $2\times10^6\,M_\odot$ at $z\sim20$ to $3\times10^8\,M_\odot$ by $z\sim9$, while the halo metallicity rises from approximately $0.03\,Z_\odot$ to $0.2\,Z_\odot$. The corresponding run without the ``\texttt{\_2}'' suffix begins its major stellar growth later, at $z\sim16$, and reaches $M_\star\simeq 2\times10^7\,M_\odot$ and $Z\sim0.1\,Z_\odot$ by $z\sim9$. The earlier enrichment of the ``\texttt{\_2}'' realization reflects the more rapid incorporation of its PBH host into the assembling main halo. The weaker early accretion feedback from the lighter PBH also allows a larger fraction of the enriched gas to remain within the halo.

For comparison, the DCBH model \citep[case C of][]{JeonSMBH2025ApJ...979..127J} exhibits a weaker correspondence between stellar-mass growth and halo metallicity. Its stellar mass increases from approximately $5\times10^6\,M_\odot$ at $z\sim11$ to $3\times10^8\,M_\odot$ by $z\sim5$, whereas its mass-weighted halo metallicity remains at only $Z\sim10^{-3.5}$--$10^{-2.8}\,Z_\odot$. By the first snapshot in which the DCBH is identified, it resides in a nearly pristine but already relatively massive halo, with $M_{\rm h,DCBH}\simeq3.7\times10^8\,M_\odot/h$ at $z\sim11$, accompanied by nearby galaxies that provide the required LW radiation. The DCBH also grows efficiently by accreting gas from the central reservoir, which, together with BH feedback, can delay the buildup of the stellar component relative to the growth of the BH and halo. Even after substantial stellar assembly, the stellar component remains a small fraction of the much larger gas reservoir, so the metals produced by star formation contribute only weakly to the halo-wide mass-weighted metallicity. The DCBH evolution pathways show more fluctuations due to partial decoupling between BH growth, stellar assembly, and chemical enrichment, than the sustained
trends seen in the overdense/enhanced-normalization PBH systems.

The simulated enrichment pathways span much of the metallicity range inferred for compact high-redshift systems~\citep[e.g.,][]{Scholtz2024A&A,Natarajan:2023UHZ1,QSO1PristineMaiolino2025, NikopoulosLRDmetal2026arXiv}. The overdense/enhanced-normalization PBH hosts progressively approach the metallicities inferred for the broader LRD population, GN-z11, and UHZ-1 as sustained star formation and enriched inflow become established. By contrast, the isolated, feedback-dominated systems experience only brief enrichment episodes and subsequently return to lower halo metallicities through metal loss and dilution, approaching the extremely metal-poor regime represented by Abell~2744-QSO1. The comparison therefore suggests that metallicity primarily traces the subsequent assembly and star-formation history of the host environment, rather than the PBH seed alone. Quantitative comparisons remain approximate because the observations generally probe gas-phase oxygen abundances in the nuclear or narrow-line-emitting region~\citep[e.g.,][]{QSO1PristineMaiolino2025, NikopoulosLRDmetal2026arXiv}, whereas Figure~\ref{fig:stellar-metallicity} shows the mass-weighted gas metallicity within the entire host halo.

\subsection{Observational Signatures\label{subsec:obs}}

\begin{figure}
    \centering
    \includegraphics[width=\linewidth]{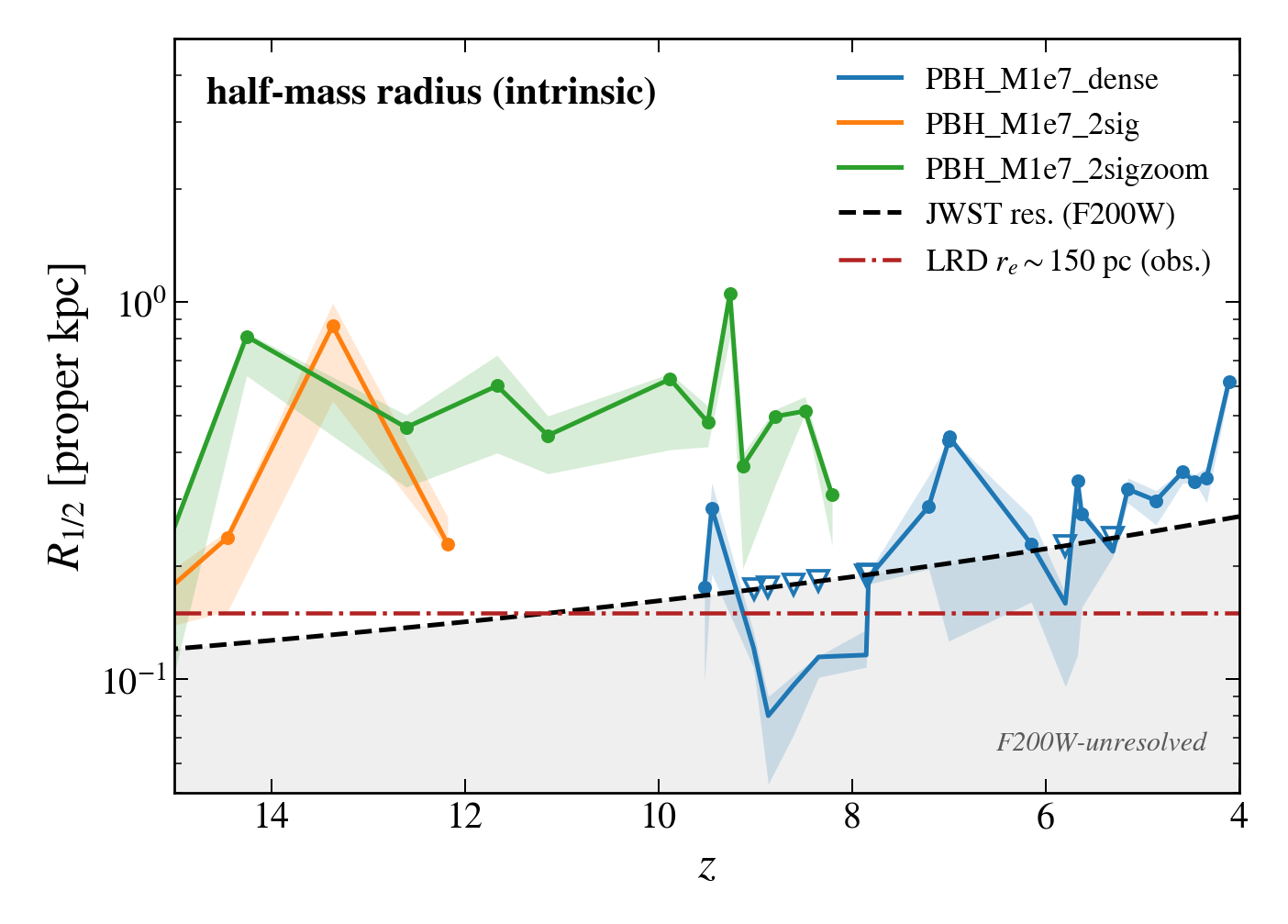}
    \caption{
Evolution of the intrinsic projected stellar half-mass radius for the $10^7\,M_\odot$ PBH simulations in overdense/enhanced-normalization environments. For each snapshot, the stellar component associated with the PBH is selected within the host-halo aperture, and the projected half-mass radius is measured along three orthogonal viewing directions. The solid curves show the median of the three projected radii, while the shaded regions span their minimum and maximum values. The black dashed curve indicates the adopted effective JWST/NIRCam F200W resolution scale~\citep{jdox_nircam_imaging}, and the open downward triangles identify snapshots whose intrinsic sizes fall below this limit. The red dash-dotted line marks a representative observed LRD effective radius of $r_e\sim150\,{\rm pc}$. The observational scales here are shown only as approximate references.
}
    \label{fig:size_evo}
\end{figure}

\begin{figure*}
    \centering
    \includegraphics[width=\linewidth]{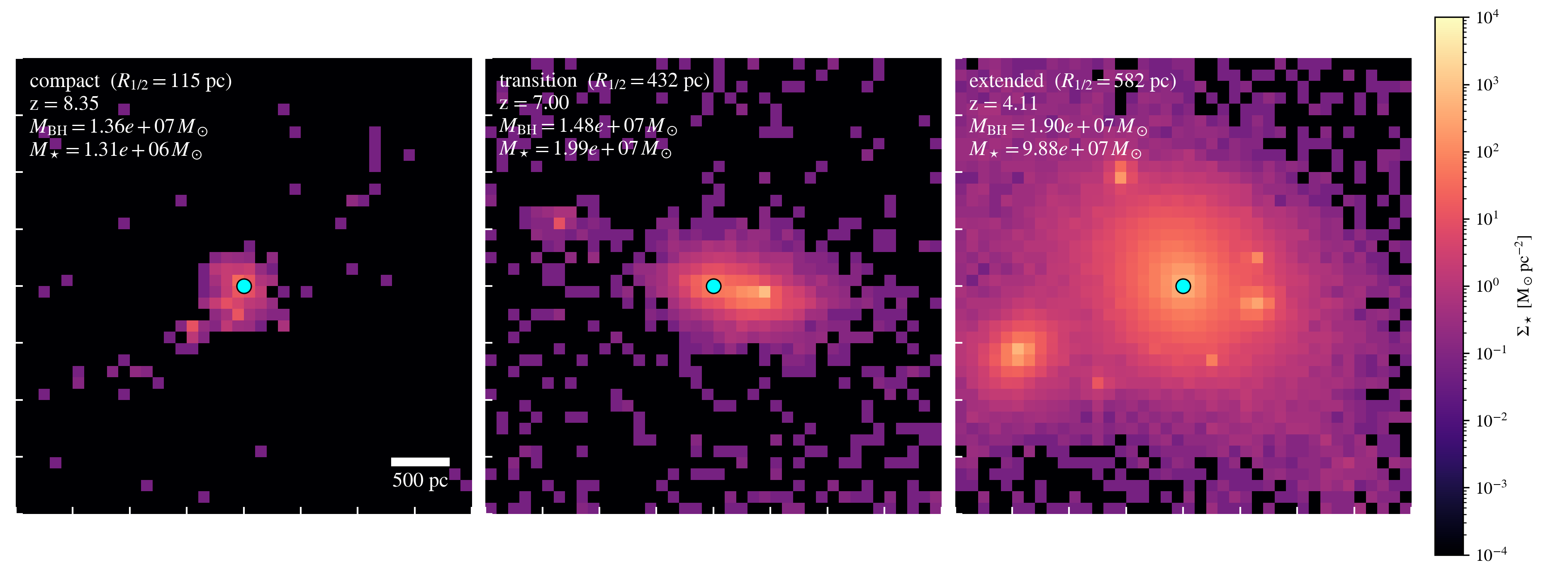}
    \caption{
Evolution of the projected stellar surface density around the PBH in the
\texttt{PBH\_M1e7\_dense} simulation at $z=8.35$, $7.00$, and $4.11$ (from left to right). Each panel covers $2\,{\rm proper\ kpc}$, adopts the same surface-density normalization and spatial binning, and is centered on the PBH, marked by the cyan circle. The stellar component evolves from a compact and irregular configuration at $z=8.35$, through a merging configuration at $z=7.00$, to an extended, centrally concentrated galaxy containing additional substructure at $z=4.11$. Over this interval, the stellar mass increases from $1.31\times10^6$ to $9.88\times10^7\,M_\odot$, whereas the BH mass grows only from $1.36\times10^7$ to $1.90\times10^7\,M_\odot$. The sequence illustrates how rapid stellar assembly, rather than substantial BH growth, moves the system away from its initially compact and BH-dominated configuration.
}
    \label{fig:2dproj}
\end{figure*}

\begin{figure*}
    \centering
    \includegraphics[width=\linewidth]{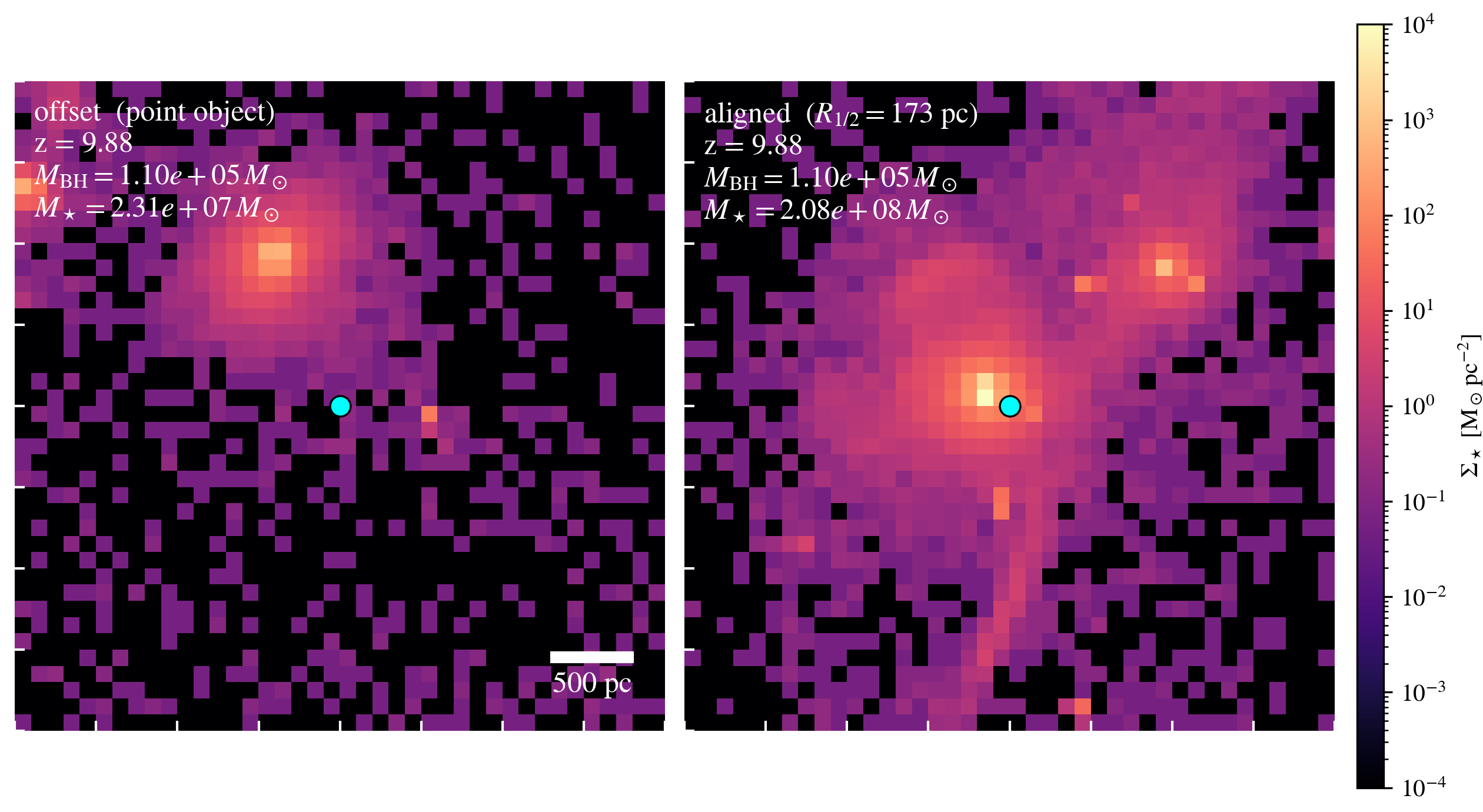}
    \caption{Projected stellar surface-density distributions around the PBHs in \texttt{PBH\_M1e5\_2sigzoom} (left) and \texttt{PBH\_M1e5\_2sigzoom\_2} (right) for the $10^5\,M_\odot$ PBH initial seed at $z=9.88$, similar to Figure~\ref{fig:2dproj}. In the offset realization, the PBH lies outside the dominant stellar concentration, despite a total stellar mass of $2.31\times10^7\,M_\odot$, therefore labeled as a point-like object. In the ``\texttt{\_2}'' realization, the PBH is associated with a prominent stellar component within a more massive and clumpy system, $M_\star=2.08\times10^8\,M_\odot$. The contrasting morphologies primarily reflect the initial PBH placement and subsequent halo assembly.
}
    \label{fig:2dprojM1e5}
\end{figure*}


In this section, we discuss the possible observational stages of PBH-seeded systems, ranging from faint or weakly fueled sources to compact LRD-like objects and more extended galaxies hosting AGNs. We use the term ``LRD-like systems'' only in an approximate evolutionary sense, referring to a stage in which a compact stellar component coexists with an accreting BH whose emission would appear as an unresolved nuclear source. 

To quantify the intrinsic stellar size, we first select star particles within the host-halo virial radius, $R_{\rm vir}$, centered on the tracked PBH. A BH-centered shrinking aperture is then applied to remove spatially distinct stellar clumps: the aperture is reduced until the mass-weighted centroid of the selected stellar component lies within $0.1R_{\rm vir}$ of the PBH. For the remaining BH-associated stellar component, we calculate the projected radius enclosing half of the stellar mass along each of three orthogonal viewing directions. Figure~\ref{fig:size_evo} shows the resulting size evolution from $z\sim15$ to $z\sim4$ for the \texttt{PBH\_M1e7\_dense}, \texttt{PBH\_M1e7\_2sig}, and \texttt{PBH\_M1e7\_2sigzoom} runs. 

In the \texttt{PBH\_M1e7\_dense} system, star formation begins at
$z\sim11$--10 in a stellar clump that is initially displaced from the PBH. After this component merges with the PBH host, continued star formation and stellar assembly build a compact BH-associated system. Its projected half-mass radius remains below or close to the nominal F200W resolution over much of $z\sim10$--5, although the intrinsic size fluctuates as new stellar clumps form and merge. At $z\lesssim5$, the host expands to $R_{1/2}\gtrsim300\,{\rm pc}$ and would become increasingly likely to appear spatially extended. By contrast, the \texttt{PBH\_M1e7\_2sig} and \texttt{PBH\_M1e7\_2sigzoom} systems generally occupy larger and more variable intrinsic sizes, $R_{1/2}\sim0.2$--$1\,{\rm kpc}$. Their earlier stellar assembly and clumpy merger histories can therefore produce resolved and spatially complex hosts at higher redshift.

To illustrate the transition away from the compact phase, we select three representative snapshots from the evolution of the \texttt{PBH\_M1e7\_dense} system and show their projected stellar surface-density distributions in Figure~\ref{fig:2dproj}. At $z\simeq8.4$, the stellar component around the PBH is compact and irregular, with $R_{1/2}\simeq110\,{\rm pc}$. At this stage, $M_\bullet\simeq1.4\times10^7\,M_\odot$ and $M_\star\simeq1.4\times10^6\,M_\odot$, corresponding to $M_\bullet/M_\star\sim10$. By $z\simeq7.0$, the stellar component has developed an elongated structure with $R_{1/2}\sim430\,{\rm pc}$, while its mass has increased to $M_\star\simeq2.0\times10^7\,M_\odot$. The corresponding mass ratio has declined to $M_\bullet/M_\star\sim0.7$, placing the system near the transition between strongly BH-dominated and host-dominated configurations.

By $z\simeq4.1$, the stellar component has developed into an extended, centrally concentrated galaxy with additional substructure on approximately kiloparsec scales. Its stellar mass reaches $M_\star\simeq9.9\times10^7\,M_\odot$, whereas the BH grows only to $M_\bullet\simeq1.9\times10^7\,M_\odot$, yielding $M_\bullet/M_\star\sim0.2$. Between the compact and extended snapshots, the stellar mass therefore increases by a factor of $\sim75$, compared with only a factor of $\sim1.4$ in BH mass. The system consequently leaves the LRD-like phase primarily through rapid assembly and spatial growth of the stellar host, making the source increasingly galaxy-like. This is qualitatively consistent with observations of LRD-like nuclei embedded in extended stellar hosts whose detectability depends on redshift and surface brightness \citep[e.g.,][]{Rinaldi2026ApJLRD}.

In the lighter-seed simulations, the spatial relation between the PBH and the stellar component introduces an additional source of diversity. Figure~\ref{fig:2dprojM1e5} compares the two $10^5\,M_\odot$ realizations at $z\simeq9.9$. In \texttt{PBH\_M1e5\_2sigzoom}, the PBH remains offset by approximately a kiloparsec from the dominant stellar cluster, which contains $M_\star\simeq2.3\times10^7\,M_\odot$. With an Eddington ratio of $\dot{M}_\bullet/\dot{M}_{\rm Edd}\sim10^{-4}$, the BH-powered emission is expected to be extremely faint and may remain below current detection limits. This geometry is qualitatively reminiscent of systems in which a weak compact source is spatially separated from the dominant stellar or line-emitting component, such as Hebe~\citep{MaiolinoHebe2026arXiv,UblerHebe2026arXiv}, although Hebe itself is more likely powered primarily by stars~\citep{JeonHebe2026ApJ,RustaHebe2026}. In the ``\texttt{\_2}'' realization, the PBH lies within a prominent stellar cluster embedded in a more massive and clumpy system, with $M_\star\simeq2.1\times10^8\,M_\odot$. Its accretion remains weak,
$\dot{M}_\bullet/\dot{M}_{\rm Edd}\sim10^{-3}$, because the BH is still displaced from the densest gas concentration. The system is therefore galaxy dominated, although its geometry is qualitatively closer to high-redshift systems such as GN-z11~\citep{Scholtz2024A&A}. These examples suggest that lighter PBH seeds may appear as either off-nuclear, weakly fueled BHs or as dynamically embedded sources, depending on their initial placement and subsequent halo assembly.

The future appearance of all the simulated systems is also expected to depend on environment. PBHs evolving in relatively isolated or average-density regions may remain associated with faint, low-surface-brightness galaxies, or fall below observational detection limits~\citep{ZhangPBHGalaxy2025ApJ, Dayal2026arXiv260420966D}. In rapidly assembling overdense regions, the same seeding channel may pass through an LRD-like phase before becoming difficult to distinguish from a more conventional high-redshift galaxy hosting an AGN. Spatially resolved JWST IFU spectroscopy, particularly when aided by strong gravitational lensing, can separate the nuclear continuum, stellar host, and line-emitting gas \citep[see, e.g.,][]{QSO1Direct2026Natur,GolubchikVENUS2026A&A, GLIMPSEBH2026ApJ}. Resolving these components would improve constraints on the stellar and dynamical masses, and the kinematics and metallicity of the surrounding gas. Combined with estimates of $M_\bullet/M_\star$ and BH accretion activity, these measurements would place stronger constraints on different seeding and evolutionary pathways.

\section{Limitations and Caveats} \label{subsec:Caveat}

Several limitations should be noted when interpreting our results. First, the simulations explore only a limited range of PBH seed masses, initial placements, and large-scale environments. Most of the analysis focuses on a fiducial seed mass of $M_{\bullet}=10^7\,M_\odot$, complemented by refined-region realizations with $M_{\bullet}=10^5\,M_\odot$. These results describe the evolution of prescribed massive PBH seeds; their cosmic abundance and formation probability are not determined by this study. If PBHs are drawn from an extended mass function, as expected in some formation scenarios~\citep[e.g.,][]{Carr2021PDU....3100755C}, their subsequent evolutionary pathways may span a wider range. The initial PBH location is also varied only in a controlled manner, by comparing an approximately average-density region with an overdense Lagrangian patch. A more complete study would need to sample the joint distribution of PBH mass and initial position in the primordial density field. This would provide a more general survey of the diverse configurations~\citep[for example triple black holes at high-$z$, see e.g.,][]{TripleBHUbler2026}, possible when PBHs composed a non-negligible fraction of dark matter.

Second, our calculations remain limited by computational cost. The full-volume simulations allow us to follow the large-scale environmental context, while the refined-region runs focus on the selected PBH-host region more efficiently. 
However, we cannot yet sample large cosmological volumes and diverse environments while also resolving individual hosts to $z\lesssim3$. The computational expense rises rapidly after substantial star formation begins, as the number of star particles increases and dense gas drives shorter time steps. The simulations should therefore be interpreted as controlled numerical experiments designed to isolate environmental trends, rather than as a statistically representative population of PBH-hosting galaxies. Establishing the relative frequency of the different evolutionary pathways will require larger volumes and a broader ensemble of initial conditions.

Third, the treatment of BH feedback is simplified. In this work, accretion feedback is implemented through thermal energy deposition in the surrounding gas, following the subgrid prescription adopted in previous studies~\citep{tremmel2017romulus}. Other feedback channels, including mechanical outflows, collimated winds, and radiation pressure, may regulate BH growth and host-galaxy evolution differently~\citep[see the review by][]{Inayoshi:2020}. These processes could affect the gas supply near the BH, the efficiency of star formation, and the spatial distribution of metals within the host galaxy. The effective thermal coupling factor $\epsilon_r$ assumed in our work also remains uncertain, and therefore should be regarded as a phenomenological parameter rather than a fully predictive quantity. Bridging the scale between the unresolved accretion flow and the galactic environment still remains a major challenge for cosmological simulations~\citep{2021ApJ...917...53A}. Future work can incorporate results from general-relativistic magnetohydrodynamics (GRMHD) simulations or connect BH growth across multiple spatial scales through refined simulations \citep[e.g.,][]{Hopkins2024OJAp....7E..18H,ChoScales2026ApJ, SuScales2026ApJ...998L..18S}.

Finally, the connection between the simulation quantities and observed high-redshift sources remains approximate. The stellar sizes reported here are intrinsic projected half-mass radii, whereas observations measure waveband-dependent, PSF-convolved half-light radii. Likewise, the simulated metallicities are averaged over an extended halo aperture, while observational estimates can be weighted more toward compact line-emitting gas near the nucleus or within bright star-forming regions depending on the adopted diagnostic lines \citep[e.g.,][]{Martinez2025,Miao2026,Moreschini2026}. A spatially extended stellar component may remain undetected because of low surface brightness or strong outshining by the BH, allowing a system to retain an LRD-like appearance even when the host is intrinsically more extended~\citep{Rinaldi2026ApJLRD}. The characteristic red continuum of LRDs is also not determined by stellar size alone, but depends on gas column density, obscuration, and radiative reprocessing. The observational benchmarks considered in this work are subject to strong selection biases, so the same behavior may not cover the broader PBH-host population. The comparisons presented here should therefore be regarded as illustrative rather than demographic. Although PBH and other heavy-seeds (DCBHs) can follow different early evolutionary tracks, observations generally probe later stages, when subsequent growth and host assembly may have weakened these differences. The combination of low metallicity and an elevated BH-to-stellar mass ratio may favor a primordial seed interpretation, but does not uniquely distinguish it from other seeding channels in an individual source. Population-level comparisons across redshift, accounting for selection effects, are needed to assess the relative support for different
seeding scenarios. A quantitative assessment will require forward modeling of the simulated stellar and BH emission using more realistic physical prescriptions, together with mock observations that account for instrumental effects and survey coverage.

\section{Summary} \label{sec:Summary}

In this work, we investigated the co-evolution of PBH seeds and their host galaxies using cosmological hydrodynamical simulations spanning different seed masses, initial placements, and large-scale environments. Our main findings are summarized as follows:

\begin{itemize}

\item
PBHs provide an early seeding channel that does not require prior baryonic collapse. Their subsequent evolution is therefore strongly regulated by the surrounding cosmological environment. Comparisons between approximately average-density regions and overdense/enhanced-normalization Lagrangian patches show that gas supply and halo assembly control both BH fueling, stellar growth and metal enrichment.

\item
The simulations produce distinct environmental pathways. In relatively isolated regions, weak gas inflow leads to intermittent BH accretion, limited star formation, and faint, metal-poor systems with elevated $M_\bullet/M_\star\gtrsim10$ by $z\sim5$. In overdense/enhanced-normalization regions, stellar assembly can begin as early as $z\sim30$--15, while continued inflow and mergers subsequently develop into more massive, metal-rich, and extended galaxies spanning values down to $M_\bullet/M_\star\sim 10^{-4}$ at $z\lesssim 8$.

\item
Host growth can dominate the evolution of the BH-to-galaxy mass ratio. In the \texttt{PBH\_M1e7\_dense} run, the BH grows only from $1.4\times10^7$ to $1.9\times10^7\,M_\odot$ between $z=8.4$ and $z=4.1$, while the associated stellar mass increases from $1.3\times10^6$ to $9.9\times10^7\,M_\odot$. The corresponding $M_\bullet/M_\star$ ratio declines from $\sim10$ to $\sim0.2$. Later galaxy assembly can therefore gradually erase the initially overmassive nature of the seed even when BH growth is non-negligible.

\item 
For the prescribed seeds studied here, some evolutionary pathways pass through an LRD-like phase, defined here by compact stellar distributions and elevated BH-to-stellar mass ratios, without establishing a match to the observed spectral properties of LRDs.
Its projected stellar half-mass radius is $R_{1/2}\simeq110\,{\rm pc}$ at $z\simeq8.4$, but subsequently grows to $\sim300\,{\rm pc}$--$1\,{\rm kpc}$ as the host assembles. The system therefore leaves the LRD-like phase primarily through rapid growth in stellar mass and spatial extent. Realizations with enhanced-normalization instead form extended and clumpy hosts at earlier times, showing that the compact phase is not universal.

\item
The lighter, $10^5\,M_\odot$ seeds remain weakly accreting and are generally galaxy dominated. Depending on their initial placement, they appear either as off-nuclear, weakly fueled BHs or as embedded but radiatively subdominant sources.

\end{itemize}

Future work should focus on identifying high-redshift PBH-seeded systems while their BH masses remain relatively modest and before continued host assembly erases the signatures of their initial conditions and drives them toward more conventional galaxy--AGN configurations. Forward modeling of their multi-wavelength and multi-messenger signals will be needed to connect the simulated evolutionary pathways to future surveys~\citep[for related work, see, e.g.,][]{DeRosa2019NewAR..8601525D, Zhou2026arXiv}. Wide-field observations with Rubin and Roman can constrain the abundance and environments of compact or off-nuclear sources \citep[e.g.,][]{Latif2025}, while ALMA can probe their cold gas and dust content \citep[e.g.,][]{DeRossi2023}. Moreover, NewAthena is expected to constrain obscured or weak BH accretion~\citep[e.g.,][]{Inayoshi2024ApJ...966..164I, LabbeLRD2025ApJ...978...92L,NewAthena2025NatAs}. Gravitational-wave observations with LISA and TianQin will provide a complementary probe of PBH binaries and their merger histories~\citep[e.g.,][]{TianQin2016CQGra..33c5010L, LISA2017arXiv170200786A,TianQin2025RPPh...88e6901L}. Combined with spatially resolved JWST follow-up, these measurements may distinguish faint offset seeds, compact BH-dominated systems, and more evolved galaxy--AGN configurations before their seeding information is lost~\citep[e.g.,][]{GolubchikVENUS2026A&A,QSO1Direct2026Natur}.

\begin{acknowledgments}
 We acknowledge fruitful discussions with Pratika Dayal, Seiji Fujimoto, and within Mike Boylan-Kolchin's group. The authors acknowledge the Texas Advanced Computing Center (TACC) for providing HPC resources under allocation AST23026. SZ and JJ gratefully acknowledge the funding from the university dissertation fellowship. BL gratefully acknowledges the funding of the Royal Society University Research Fellowship and the Deutsche Forschungsgemeinschaft (DFG, German Research Foundation) under Germany's Excellence Strategy EXC 2181/1 - 390900948 (the Heidelberg STRUCTURES Excellence Cluster). P.N. acknowledges support from the Gordon and Betty Moore Foundation and the John Templeton Foundation, which fund the Black Hole Initiative (BHI) at Harvard University, where she serves as a PI. P.N. also acknowledges support from STScI/NASA via grant JWST-GO-03293024. We used ChatGPT and Claude for language editing and stylistic suggestions. The authors reviewed and verified all scientific content, interpretations, calculations, and references.
\end{acknowledgments}

\begin{contribution}

SZ is responsible for implementing the numerical analysis and writing up the manuscript; JJ has provided relevant data for the DCBH simulations; BL and VB have helped with overall narratives, science directions, and critical edits; PN has helped with science pertaining to PBH model and overall narratives.


\end{contribution}

%
\vspace{5mm}
\facilities{Lonestar6 and Stampede3 (TACC)}


\software{GIZMO~\citep{Hopkins2015MNRAS.450...53H}, astropy~\citep{2013A&A...558A..33A,2018AJ....156..123A, 2022ApJ...935..167A},  
          Colossus~\citep{Diemer2018ApJCOLOSSUS}, PHANTOM~\citep{zhang2025PHANTOM}, NumPy~\citep{Harris20}, SciPy~\citep{SciPy2020}, Matplotlib~\citep{Hunter2007}, ChatGPT~\citep{openai_chatgpt_2026}, Claude~\citep{anthropic_claude_2026}
          }


\appendix
\vspace{-10pt}
\section{Abundance and clustering in the single-seed regime}\label{Appendix:A}

Our simulations follow a single massive PBH within the simulated region. For a given cosmic PBH abundance and clustering strength, we estimate the probability $P_{\rm single}$ that a selected seed has no additional PBH in this region. We then assess whether abundances that yield a high $P_{\rm single}$ can still provide an astrophysically relevant population of early SMBH seeds.

For a monochromatic PBH population of mass $M_\bullet$, let $f_{\rm PBH}$ denote its cosmic mass fraction relative to dark matter, evaluated at $z_{\rm ini}=3400$. Its mean comoving number density is
\begin{equation}
 \bar n_{\rm PBH}
 =\frac{f_{\rm PBH}\bar\rho_{\rm DM,0}}{M_\bullet}.
 \label{eq:pbh_number_density}
\end{equation}
Placing one seed in a selected simulation region does not imply a cosmic mean occupation of one PBH per box.

Excluding the central PBH, we write the conditional companion density as $\bar n_{\rm PBH}[1+\xi(\mathbf x)]$. Here $\xi$ includes conditioning on the selected matter environment; without this selection, it is the usual PBH pair correlation~\citep{Desjacques2018PhRvD}. The expected companion count is
\begin{equation}
 \lambda\equiv\langle N_{\rm comp}\rangle
 =\bar n_{\rm PBH}V(1+\bar\xi),
 \qquad
 \bar\xi=\frac{1}{V}\int_V\xi(\mathbf x)\,d^3x.
 \label{eq:lambda}
\end{equation}
Here $N_{\rm comp}$ denotes the number of additional PBHs within the simulated region $V$. We define the single-seed probability as

\begin{equation}
 P_{\rm single}\equiv
 P(N_{\rm comp}=0\mid
 \text{central PBH, selected environment}).
\end{equation}

The complementary probability describes a region containing at least one additional PBH. Since $N_{\rm comp}$ is a nonnegative integer with mean $\lambda$,

\begin{equation}
 1-P_{\rm single}
 =P(N_{\rm comp}\geq1\mid
 \text{central PBH, selected environment})
 \leq\min(1,\lambda).
 \label{eq:single_probability}
\end{equation}

This bound does not assume Poisson statistics.
We adopt $P_{\rm single}\geq0.9$ as a benchmark for the applicability of our single-seed approximation. For our $L=2\,{\rm cMpc}/h$ box, $V=L^3=25.83\,{\rm cMpc}^{3}$ and
$\bar\rho_{\rm DM,0}=3.33\times10^{10}\,
M_\odot\,{\rm cMpc}^{-3}$, a sufficient condition $\lambda\leq0.1$ gives
\begin{equation}
 f_{\rm PBH}\leq
 \frac{1.16\times10^{-8}}{1+\bar\xi}
 \left(\frac{M_\bullet}{10^5M_\odot}\right).
 \label{eq:single_pbh_bound}
\end{equation}

Thus, stronger clustering requires a lower cosmic abundance to satisfy the same isolation criterion, as shown in Table~\ref{Table:PBHParam}. This is a sufficient condition for the single-seed approximation, not an observational exclusion limit. Failing this condition does not establish that isolated seeds are absent.

\begin{table*}[htb!]
 \centering
 \caption{Sufficient upper limits on the cosmic PBH fraction
 for $\lambda\leq0.1$, for illustrative conditional clustering
 enhancements. These limits describe the applicability of
 the single-seed approximation.}
 \begin{tabular}{ccc}
 \hline
 $1+\bar\xi$
 & Upper limit on $f_{\rm PBH}$ ($M_\bullet=10^5M_\odot$)
 & Upper limit on $f_{\rm PBH}$ ($M_\bullet=10^7M_\odot$)\\
 \hline\hline
 $1$    & $1.16\times10^{-8}$  & $1.16\times10^{-6}$\\
 $10^2$ & $1.16\times10^{-10}$ & $1.16\times10^{-8}$\\
 $10^3$ & $1.16\times10^{-11}$ & $1.16\times10^{-9}$\\
 \hline
 \end{tabular}
 \label{Table:PBHParam}
\end{table*}

These small mass fractions can still provide an astrophysically relevant seed population. For example, $\bar n_{\rm PBH}=10^{-5}\,{\rm cMpc}^{-3}$ corresponds to $f_{\rm PBH}\simeq3.0\times10^{-11}$ and $3.0\times10^{-9}$ for $M_\bullet=10^5$ and $10^7\,M_\odot$, respectively. With a strong clustering enhancement, $1+\bar\xi=100$, this gives $\lambda=0.026$ and a companion probability below $2.6\%$. Even for $1+\bar\xi=10^3$, the ten-percent criterion allows mean seed densities up to $3.9\times10^{-6}\,{\rm cMpc}^{-3}$.

For comparison, selected high-redshift LRD samples have number densities of $\sim10^{-5}$--$10^{-4}\,{\rm cMpc}^{-3}$~\citep{KokorevLRD2024ApJ...968...38K,Ma2026LRDApJ}. The example population with $\bar n_{\rm PBH}=10^{-5}\,{\rm cMpc}^{-3}$ is therefore comparable to the lower end of this range. This illustrative population could contribute to a subset of LRDs under their accreting-BH interpretation. The observable fraction depends on the host environment, subsequent growth, and activity; it is not determined by the companion-count calculation.

The clustering enhancement is a model-dependent input. Explicit calculations in the Affleck--Dine scenario show that the PBH correlation function can vary substantially even at fixed abundance~\citep{Kasai2025PhRvD}. Its strength depends on spatial correlations and long--short mode coupling, rather than on one-point non-Gaussianity alone~\citep{Suyama2019PTEP}. We therefore use the values in Table~\ref{Table:PBHParam} as illustrative inputs. A specific formation model must predict both the abundance and the correlation function, which must then be averaged over the relevant volume at
$z_{\rm ini}$ to give $\bar{\xi}$.

\section{Effect of Varying Black Hole Feedback Efficiency}\label{Appendix:B}

\begin{table*}[htb!]
    \centering
    \caption{Similar to Table~\ref{Table:SimParam}, we summarize the parameters of the additional simulations used to examine the dependence on the BH feedback coupling efficiency $\epsilon_r$. All three runs share the same refined-region initial conditions and differ only in $\epsilon_r$, so we eliminated columns with the same parameter. We additionally list the final BH mass $M_\bullet$, redshift of the onset of star formation $z_\star$, total stellar mass $M_\star$, and mean metallicity $Z/Z_\odot$ within the PBH host halo, all measured at $z\simeq 9.5$.}
    \begin{tabular}{ccccccccc}
    \hline
        Run &$z_{\rm ini}$  & $z_{\rm final}$&  $\epsilon_{\rm r}$ &$\langle f_{\rm edd}\rangle$ & $M_\bullet (z\simeq9.5) [\,M_{\odot}]$& $z_\star$ & $M_\star(z\simeq9.5)[\,M_{\odot}]$  & $Z(z\simeq9.5)/Z_\odot$ \\
    \hline
    \texttt{PBH\_M1e7\_2sigzoom\_fd05}   & 1100 & 9.5 &    0.05 & 0.0080 &$1.12 \times 10^7$ & 24.5 & $1.11\times 10^8$ & 0.12  \\
    \texttt{PBH\_M1e7\_2sigzoom}   & 1100 & 8.1 & 0.005 & 0.029 & $1.72 \times 10^7$ & 27.4 & $5.26\times 10^7$  & 0.051  \\
    \texttt{PBH\_M1e7\_2sigzoom\_fd001}   & 1100 & 9.1 &    0.001 & 0.069 & $4.95 \times 10^7$ & 31.4 & $6.61\times 10^7$ & 0.061 \\
    \hline \hline
    \end{tabular}
    \label{Table:SimParam_feedback}
\end{table*}

\begin{figure*}[htb!]
    \centering
    \includegraphics[width=\linewidth]{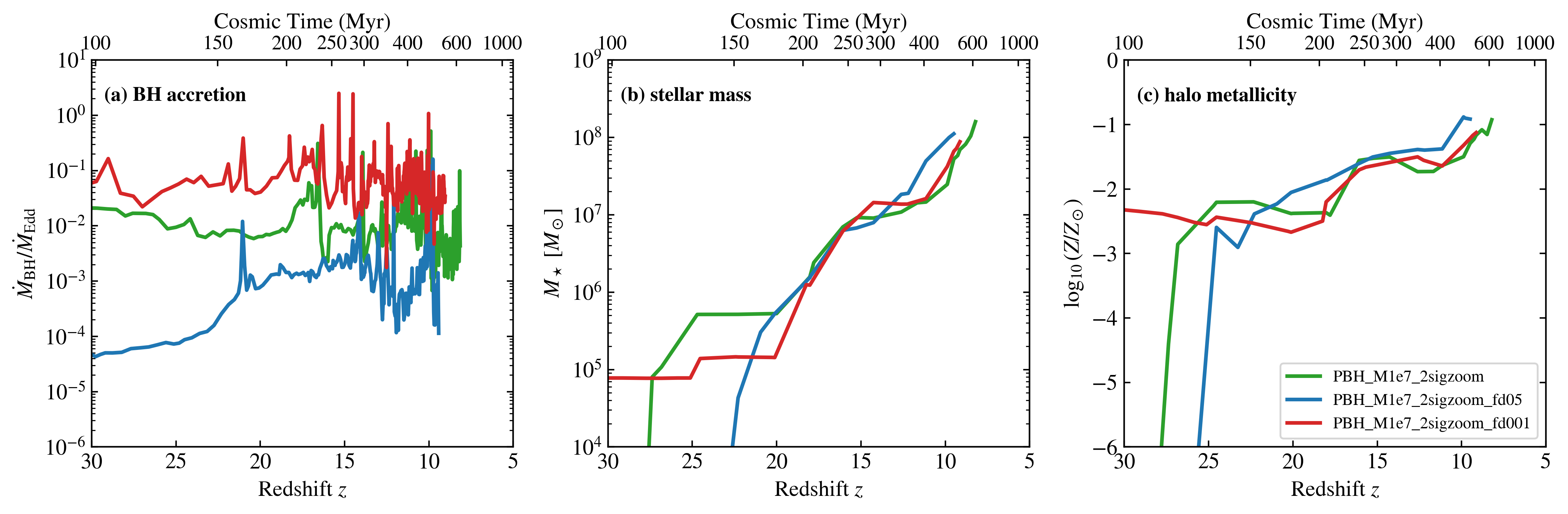}
    \caption{Dependence of the overdense/enhanced-normalization PBH evolution on the adopted BH feedback coupling efficiency. From left to right, we show the Eddington ratio, total stellar mass, and mean metallicity within the PBH host halo as functions of redshift. All three simulations use the same refined-region initial conditions and differ only in the feedback coupling efficiency: the green curve denotes the fiducial model with $\epsilon_r=0.005$ (\texttt{PBH\_M1e7\_2sigzoom}), while the blue and red curves correspond to $\epsilon_r=0.05$ (\texttt{PBH\_M1e7\_2sigzoom\_fd05}) and $\epsilon_r=0.001$ (\texttt{PBH\_M1e7\_2sigzoom\_fd001}), respectively. Lower coupling efficiencies permit substantially higher BH accretion and an earlier onset of star formation. At later times, however, the stellar masses and metallicities of the three hosts become considerably more similar, indicating a weaker dependence of the integrated host properties on the adopted BH feedback efficiency.}
    \label{fig:feedback_test}
\end{figure*}

To examine the sensitivity of the host evolution to the adopted BH feedback prescription in overdense/enhanced-normalization realizations, we perform two additional runs with the same initial conditions as the \texttt{PBH\_M1e7\_2sigzoom} run, but with feedback coupling efficiencies $\epsilon_r=0.05$ and $0.001$, bracketing our fiducial value of $\epsilon_r=0.005$. The resulting evolution is shown in Figure~\ref{fig:feedback_test}, with the main properties summarized in Table~\ref{Table:SimParam_feedback}. Consistent with our previous findings~\citep{ZhangPBHGalaxy2025ApJ, Zhang2026PBHApJL}, BH growth is strongly sensitive to the adopted coupling efficiency. Decreasing $\epsilon_r$ from $0.05$ to $0.001$ raises the time-averaged Eddington ratio from $\langle f_{\rm Edd}\rangle\simeq0.008$ to $0.069$, producing a correspondingly larger final BH mass. 

The onset of star formation also depends on the feedback strength: weaker coupling allows the surrounding dense gas to cool and collapse earlier, shifting the onset of star formation from $z\simeq24.5$ in the strongest-feedback case to $z\simeq31.4$ in the weakest-feedback run. 
The subsequent evolution, however, shows a weaker and non-monotonic dependence on $\epsilon_r$. Stronger feedback delays the initial collapse and star formation, whereas in the weak-feedback case, the more rapidly growing BH competes more efficiently with star formation for the dense central gas. Consequently, weaker BH feedback does not simply translate into a larger stellar mass. Once larger-scale halo assembly becomes established, continued cosmological gas supply increasingly regulates the assembly of the host, and the three runs reach comparable stellar masses and metallicities at $z\lesssim 18$. 
Thus, the feedback tests preserve the qualitative trend of stellar mass buildup and chemical enrichment, with all three runs reaching $M_\star>M_\bullet$ by $z\simeq9.5$. However, the timing of this evolution and the mass ratios, accretion rates, and metallicities at a given redshift remain sensitive to the feedback efficiency, affecting comparisons with individual observed sources.


\bibliography{Main}{}
\bibliographystyle{aasjournalv7}



\end{document}